\documentclass{article}

\usepackage{lipsum}
\usepackage{amsfonts}
\usepackage{graphicx}
\usepackage{epstopdf}
\usepackage{algorithmic}
\usepackage{amsopn}

\usepackage{graphicx} 
\usepackage{xcolor}
\usepackage{amsfonts}
\usepackage{amsmath}
\usepackage{amssymb}
\usepackage{amstext}
\usepackage{amsthm}
\usepackage{bm}
\usepackage{url}
\usepackage{subcaption}
\usepackage{stmaryrd}
\usepackage{multirow}
\usepackage{esint} 
\usepackage[colorlinks=true,urlcolor=blue,linkcolor=blue,citecolor=blue]{hyperref}
\usepackage[group-separator={,},group-minimum-digits={3}]{siunitx}

\usepackage[capitalise,noabbrev]{cleveref}
\crefformat{equation}{(#2#1#3)}
\numberwithin{equation}{section}
\numberwithin{figure}{section}
\usepackage[normalem]{ulem}

\def\tCRALS{\text{CR}_\text{ALS}}

\def\tCRALSS{\text{CR}_\text{ALS-skew}}
\def\tCRSVD{\text{CR}_\text{SVD}}
\def\tfluc{\text{fluc}}

\newcommand{\efluc}{\mathrm{E}(t)_{\tfluc}}
\newcommand{\energy}{\mathrm{E}(t)}
\newcommand{\email}[1]{\texttt{#1}}

\ifpdf
  \DeclareGraphicsExtensions{.eps,.pdf,.png,.jpg}
\else
  \DeclareGraphicsExtensions{.eps}
\fi

\newtheorem{remark}{Remark}

\crefname{hypothesis}{Hypothesis}{Hypotheses}
\crefname{fact}{Fact}{Facts}

\title{Accelerating Galerkin Reduced-Order Models for Turbulent Flows with Tensor Decomposition}

\author{Ping-Hsuan Tsai\thanks{Corresponding Author. Department of Mathematics, Virginia Tech, Blacksburg, VA 24061
  (\email{pinghsuan@vt.edu}).} \and Paul Fischer\thanks{Department of Mechanical Science \& Engineering and Siebel School of Computing and Data Science, University of Illinois at Urbana-Champaign, Urbana, IL, 61801, (\email{fischerp@illinois.edu})} \and Edgar Solomonik\thanks{Siebel School of Computing and Data Science, University of Illinois at Urbana-Champaign, Urbana, IL, 61801, (\email{solomon2@illinois.edu})}
  }

\def\bX{{\bf X}}

\def\cN{{\cal N}}

\def\scriptO{{{\it O}\kern -.42em {\it `}\kern + .20em}}
\def\RR{{{\rm l}\kern - .15em {\rm R} }}
\def\PP{{{\rm l}\kern - .15em {\rm P} }}
\def\L2{{{\sf L}^2}}
\def\H1{{{\sf H}^1}}
\def\PN2{{\PP_{N}-\PP_{N-2}}}

\def\complex{{{\rm C} \kern - .53em {\rm l} \kern + .38em}}

\def\a1{{ | \lambda_{\min} |}}

\def\l1{{   \lambda_{\min}  }}

\def\bn{{{\bf n}}}

\def\bu0{{\underline {\bf 0}}}

\def\bu{{\bf u}}

\def\bv{{\bf v}}
\def\bx{{\bf x}}

\def\Oh{{\hat \Omega}}

\def\ua{{\underline a}}

\def\uu{{\underline u}}

\def\u0{{\underline 0}}
\def\1u{{\underline 1}}

\newcommand{\pp}[2]{\frac{\partial #1}{\partial #2} }

\def\tr{{\tilde r}}

\def\cN{{\cal N}}

\def\mX{{\mathcal{X}}}

\def\mC{{\mathcal{C}}}

\def\bbu{{\overline{\bf u}}}

\def\mC{\mathcal{C}}
\def\mO{\mathcal{O}}
\def\mX{\mathcal{X}}

\def\tmC{\tilde{\mC}}
\def\wmC{{\widehat{\mathcal{C}}}}

\def\bu{{\bm u}}
\def\bv{{\bm v}}

\def\tr{\text{r}}
\def\relres{r_\text{rel}}
\newcommand{\bphi}{\boldsymbol{\varphi}}

\usepackage{parskip} 
\usepackage{lmodern} 
\usepackage{etoolbox}   
\newcommand{\Keywords}{%
  \vspace{1ex}
  \noindent\textbf{Keywords.}%
}

\newcommand{\MSCcodes}{%
  \vspace{1ex}
  \noindent\textbf{Mathematics Subject Classification (MSC2020).}%
}

\begin{document}

\maketitle

\begin{abstract}
   Galerkin-based reduced-order models (G-ROMs) 
   offer efficient and
   accurate approximations for 
   laminar flows but require hundreds to thousands of modes $N$  
   to capture the complex dynamics of turbulent flows. This makes standard G-ROMs computationally expensive due to the third-order advection tensor contraction, 
   requiring the storage of $N^3$ entries 
   and the computation of $2N^3$ 
   operations per timestep. As a result, 
   such ROMs are impractical for realistic 
   applications like turbulent flow control. 
   In this work, we 
   consider problems that demand large $N$ values for accurate G-ROMs and 
   propose a novel
   approach that accelerates G-ROMs by utilizing the CANDECOMP/PARAFAC (CP) tensor decomposition to approximate the advection tensor as a sum of $R$ rank-1 tensors. We also leverage the partial skew-symmetry property of the advection tensor and derive two conditions for the CP decomposition to preserve this property.
   Moreover, we investigate the low-rank structure of the advection tensor using singular value decomposition (SVD) and compare the performance of G-ROMs accelerated by CP (CPD-ROM) and SVD (SVD-ROM). Demonstrated on problems from 2D periodic to 3D turbulent flows, the CPD-ROM achieves at least a $10$-fold speedup and a $16.7$-fold reduction in nonlinear term evaluation costs compared to the standard G-ROM. The skew-symmetry preserving CPD-ROM demonstrates 
   improved stability in both the reproduction and predictive regimes, and enables the use of smaller rank $R$. Singular value analysis reveals a persistent low-rank structure in the $H^1_0$-based advection tensor, and CP decomposition achieves at least an order of magnitude higher compression ratio than SVD.
\end{abstract}

\Keywords\ Reduced-order modeling, Galerkin projection, turbulent flows, tensor decomposition, alternating least squares, skew-symmetric tensors.

\MSCcodes\ 65L05, 15A69, 15A23, 76D05, 76F99.

\section{Introduction}
\label{section:introduction}

Reduced order models (ROMs) are computational models that leverage data to capture the essential dynamics of complex systems such as fluid flow in a low-dimensional space whose dimension, $N$, is orders of magnitude
lower than the dimension of full order models (FOMs). 
FOMs are typically obtained through classical numerical discretizations, such as finite element, finite difference, or finite volume methods. In the numerical simulation of fluid flows, Galerkin ROMs
(G-ROMs), which use data-driven bases within a Galerkin framework, have provided
efficient and accurate approximations of laminar flows, such as the
two-dimensional flow past a circular cylinder at low Reynolds
numbers~\cite{hesthaven2015certified,quarteroni2015reduced}.

For nonlinear problems, ROMs are no longer efficient because the evaluation of the nonlinear term depends on the FOM degrees of freedom. To address this, a variety of hyper-reduction techniques have been developed, including the empirical interpolation method (EIM) \cite{barrault2004empirical}, discrete empirical
interpolation method (DEIM)
\cite{chaturantabut2010nonlinear},
missing point estimation \cite{astrid2008missing}, gappy POD
\cite{carlberg2013gnat}, S-OPT \cite{lauzon2024s}, and others. These methods enable efficient evaluation of nonlinear terms and make parametric model order reduction (pMOR) feasible for nonlinear systems. In the case of the Navier–-Stokes equations (NSE), the advection operator is nonlinear but polynomial, which allows 
its evaluation in the ROM to be independent of the FOM degrees of freedom.
In particular, a third-order reduced advection tensor can be precomputed during the offline stage. During the online stage, the ROM advection term can then be evaluated through tensor contraction, requiring only $2N^3$ operations per timestep.  
While this cost is acceptable for low-dimensional ROMs, it becomes prohibitive for turbulent flows, which typically require hundreds or thousands of basis functions~\cite[Table II]{ahmed2021closures},\cite{tsai2025time} to 
capture the complex dynamics of the turbulent flow. Although the resulting ROM is still relatively
low-dimensional compared to the FOM, its computational cost becomes prohibitive.
The 3rd-order tensor requires storage of $N^3$ entries with a corresponding work
of $2N^3$ operations per timestep. While $N=100$, with a cost of two million
operations per step and a million words in memory, may be tolerable, $N=400$
with a cost of $128$ million operations and $64$ million words, making such ROMs impractical for realistic applications like control of turbulent flows.

Although several efforts have applied hyper-reduction techniques to NSE, their focus is different from reducing the cost of the advection term. For example, EIM has been successfully combined with iterative solvers to efficiently compute steady solutions of the incompressible NSE \cite{elman2017numerical}. For the unsteady NSE, DEIM is used to handle the strong nonlinearities in non-hydrostatic models \cite{xiao2014non}, and other hyper-reduction strategies have been developed to conserve kinetic energy and momentum \cite{klein2024energy}. In contrast, applying hyper-reduction to reduce the cost of evaluating the advection term 
has not received much attention, primarily because, as discussed above, the nonlinearity is polynomial in nature and can be precomputed offline. Moreover, in many studies, the reduced dimension $N$ is small (typically $N<
100$), making the cost of tensor contraction manageable. 

In this work, we focus on the scenario where a large number of basis functions ($N>100$), 
is required in G-ROM to accurately capture the complex dynamics 
and investigate the potential of the tensor decomposition for reducing the tensor
contraction cost. 
Tensor decomposition has been widely used as a low-rank approximation for
reducing the cost at the FOM level. 
For example, in \cite{dolgov2017low}, a low-rank tensor decomposition algorithm has been developed for the numerical solution of a distributed optimal control problem constrained by the two-dimensional time-dependent NSE with a stochastic inflow.
Other decompositions, such as the hierarchical Tucker decomposition and the tensor train, have also been 
applied to other types of equations, such as the Vlasov equation \cite{einkemmer2020low}, the Fokker–-Planck equation \cite{sun2014numerical}, and the Boltzmann equation \cite{chikitkin2021numerical}.
The dynamical low-rank approximation has been used to compute low-rank approximations to time-dependent large data matrices or to solutions of large matrix differential equations \cite{koch2010dynamical,nonnenmacher2008dynamical}. In the context of reduced-order modeling, tensor decomposition has been used for various purposes, such as extracting multiple space-time basis vectors from each training simulation \cite{choi2019space}. In \cite{shimizu2021windowed}, the author considered the windowed space-time least-squares Petrov--Galerkin method for model reduction of nonlinear parameterized dynamical systems and proposed constructing space-time bases using tensor decompositions for each window. More recently, tensor decomposition was employed in \cite{olshanskii2025approximating} to develop a reduced-order model for the parametric unsteady NSE.

In this paper, we propose a novel approach to accelerate G-ROMs by utilizing 
the 
CANDECOMP\slash{}PARAFAC (CP)
tensor decomposition to approximate the ROM advection tensor as a sum of $R$ rank-1 tensors (CPD-ROM). 
We analyze the partial skew-symmetry of the ROM advection tensor and derive conditions under which the CP decomposition preserves this property.
The CPD-ROM is investigated in several test problems from 2D periodic flow to 3D turbulent flows.
In addition, we investigate the impact of approximating the full tensor versus the core tensor in the CPD-ROM, and whether preserving skew-symmetry in the CP decomposition further improves its performance.
Moreover, we investigate the low-rank structure of the advection tensor constructed using $L^2$-POD and $H^1_0$-POD basis functions with singular value decomposition (SVD) and compare the performance of the CPD-ROM with the SVD-ROM, where the advection tensor is approximated using SVD, in terms of memory savings and cost reduction. 

The rest of the paper is organized as follows: In Section \ref{sec:background}, we provide the backgrounds for the full-order model (FOM), the Galerkin reduced-order model (G-ROM), and the CANDECOMP/PARAFAC (CP) tensor decomposition for general and partially symmetric tensors. 
In Section \ref{section:acc-cpd-rom}, we present the CPD-ROM, in which the ROM advection tensor is approximated using the CP decomposition. We also demonstrate that the tensor is partially skew-symmetric under divergence-free and certain boundary conditions and introduce a CP decomposition that preserves this skew-symmetric property. In addition, we show that the ROM tensor has an underlying CP structure.
In Section \ref{section:numerical-results}, we present numerical results for the investigations outlined above. In Section \ref{sec:conclusions}, we present the conclusions of our numerical investigation and outline directions for future research.
\section{Background} 
    \label{sec:background}

In this section, we provide the necessary background for the paper. In Section
\ref{subsection:fom}, we briefly introduce the FOM and
suggest \cite{tsai2022parametric} for a detailed review of it.  In Section
\ref{subsection:g-rom}, we present the Galerkin reduced-order model (G-ROM).
In Sections \ref{subsection:cpd}-\ref{subsection:cpd_sym}, we introduce the
CP decomposition with the alternating least squares method
(ALS) for general and partially symmetric tensors.

\subsection{Full-order model (FOM)} \label{subsection:fom}
The governing equations are the incompressible Navier--Stokes equations (NSE):
\begin{equation} \label{equation:nse}
  \frac{\partial \bu}{\partial t} + (\bu \cdot \nabla)\bu 
    = -\nabla p + \frac{1}{\rm Re} \nabla^2 \bu, \quad 
   \nabla \cdot \bu = 0,
\end{equation}
where $\bu$ represents the velocity subject to appropriate Dirichlet or Neumann boundary conditions, $p$ denotes the pressure,
and $\rm Re$ is the Reynolds number. 

We employ the spectral element method (SEM) for the spatial discretization of (\ref{equation:nse}).
The $P_{q}$--$P_{q-2}$ velocity-pressure coupling \cite{maday1987well} is considered where the velocity $\bu$ is represented as a tensor-product Lagrange polynomial of degree $q$ in the reference element $\Oh := [-1,1]^2$
while the pressure $p$ is of degree $q-2$,
resulting a velocity space $\bX^{\cN}$ for approximating the velocity and a pressure space $Y^{\bar{\cN}}$ for approximating the pressure, where the finite dimensions $\cN$ and $\bar{\cN}$ are the global numbers of spectral element degrees of freedom in the corresponding spectral element spaces.

The semi-discrete weak form of
(\ref{equation:nse}) reads: {\em Find $(\bbu,~p) \in
(\bX^{\cN}, Y^{\bar{\cN}})$ such that, for all $(\bv,~q) \in (\bX^{\cN}_0,Y^{\bar{\cN}}_0)$,}
\begin{align} \label{eq:var1}
   \left( \bv,\pp{\bbu}{t} \right)  + \biggl( (\bbu \cdot \nabla) \bbu, \bv \biggr) & = \left( \nabla \cdot
   \bv , p \right) - \frac{1}{\rm Re} \, (\nabla \bv, \nabla \bbu),
   \\[1.6ex] \label{eq:var2}
   - \left(q, \nabla \cdot \bbu  \right)  &= 0,
\end{align}
where $(\cdot, \cdot)$ denotes as the $L^2$ inner product. The diffusive and the
pressure terms have different signs due to the application of
integration by parts and the divergence theorem. Following \cite{fischer2017recent}, a semi-implicit scheme BDF$k$/EXT$k$ is
employed for time discretization resulting in a fully discretized linear
unsteady Stokes system.  The detailed derivation of the FOM is referred to
\cite{tsai2022parametric}. 

\subsection{Galerkin reduced-order model (G-ROM)} \label{subsection:g-rom}

We employ the standard proper orthogonal decomposition (POD) procedure
\cite{berkooz1993proper} to construct the reduced basis
function. We begin by collecting a set of velocity snapshots $\{\bu^k :=
\bu(\bx,t^k)-\bphi_0\}^K_{k=1}$, which correspond to the FOM solutions at
well-separated timepoints $t^k$, with the subtraction of the zeroth mode $\bphi_0$.
In this study, the zeroth mode $\bphi_0$ is set to be the time-averaged
velocity field in the time interval in which the snapshots were collected.
The Gramian matrix is then formed using the $L^2/H^1_0$ inner product and
the first $N$ POD basis functions $\{\bphi_i\}^N_{i=1}$ are constructed from the
first $N$ eigenmodes of the Gramian. 
The G-ROM is constructed by inserting the ROM basis expansion
\begin{equation} \label{equation:romu}
   \bu_\tr(\bx) = \bphi_0(\bx) + \sum_{j=1}^N u_{\tr,j} \bphi_j(\bx)
\end{equation}
into 
the weak form of (\ref{equation:nse}):
{\em Find $\bu_\tr$ 
such that, for all $\bv \in \bX^N_0$,}
\begin{eqnarray}
    && 
    \left(
        \bv_{i}, \frac{\partial \bu_\tr}{\partial t} 
    \right)
    + \frac{1}{\rm Re} \, 
    \left( 
        \nabla \bv_{i},
        \nabla \bu_\tr 
    \right)
    + \biggl( 
        \bv_{i},
        (\bu_\tr \cdot \nabla) \bu_\tr
    \biggr)
    = 0,  
    \label{equation:gromu}
\end{eqnarray}
where $\bX^N_0 := \text{span} \{\bphi_i\}^N_{i=1}$ is the ROM space.

\begin{remark}
We note that, in the case of fixed geometries, the divergence and
pressure terms drop out of (\ref{equation:gromu}) since the 
ROM basis is weakly divergence-free. For ROMs that include the pressure approximation, see, e.g.,~\cite{decaria2020artificial,hesthaven2015certified,noack2005need,quarteroni2015reduced}. 
\end{remark}

With (\ref{equation:gromu}), a system of differential equations in the coefficients with respect to the POD bases $u_{\tr,j}$ are derived:
\begin{align}
	B \frac{d \uu_{\tr}}{dt} & =
	-\mC (\bar{\uu}_{\tr}) \bar{\uu}_{\tr} - \frac{1}{\rm Re}
	 A \bar{\uu}_{\tr}, \label{equation:nse_ode}
\end{align}
where $\uu_\tr \in \mathbb{R}^N$ is the vector consists of POD coefficients $\{ u_{r,j}\}^N_{j=1}$ and
$\bar{\uu}_{\tr} \in \mathbb{R}^{N+1}$ is the augmented vector that includes the zeroth mode's coefficient. $A$, $B$, and $\mC$ represent the 
reduced stiffness, mass, and advection
operators, respectively, with entries
\begin{align} 
A_{ij} =\int_{\Omega}\nabla \bphi_i :\nabla \bphi_j\, dV, 
\;\; B_{ij} =\int_{\Omega}\bphi_i\cdot\bphi_j\, dV, 
\;\; \label{equation:Cu} \mC_{ikj} =\int_{\Omega}\bphi_i\cdot(\bphi_k\cdot\nabla)\bphi_j\,dV.
\end{align}

For the temporal discretization of (\ref{equation:nse_ode}), a semi-implicit scheme with $k$th-order backward differencing (BDF$k$) and $k$th-order extrapolation (EXT$k$) is considered. 
The fully discretized reduced system at time $t^l$ is 
\begin{align}
	\left(\frac{\beta_0}{\Delta t} B + \frac{1}{\rm Re} A \right) \, \uu_{\tr}^{l+1} & = - \sum^k_{i=1} \alpha_i \left[ \mC ({\uu}^{l-i}) {\uu}^{l-i} + (C_1 + C_2) \bar{\uu}^{l-i} - \underline{c}_{0} \right] \nonumber \\ & - B \sum^k_{i=1}\frac{\beta_i}{\Delta t} \uu^{l-i} - \frac{1}{\rm Re} \underline{a}_0,
\label{equation:nse_d1}
\end{align}
where
\begin{align}
c_{0,i} & = \int_{\Omega}\bphi_i\cdot(\bphi_0\cdot\nabla)\bphi_0\,dV,\quad
a_{0,i} = \int_{\Omega}\nabla \bphi_i : \nabla \bphi_0 \,dV, \\ 
C_{1,ij} & = \int_{\Omega}\bphi_i\cdot(\bphi_0\cdot\nabla)\bphi_j\,dV,\quad
C_{2,ik} = \int_{\Omega}\bphi_i\cdot(\bphi_k\cdot\nabla)\bphi_0\,dV, \label{equation:zeroth_contribution}
\end{align}
for all $i=1,\ldots,N$ and $j,~k=0,\ldots,N$.

\begin{remark}
The computational cost of solving the reduced system (\ref{equation:nse_d1}) is dominated by the application of the
rank-3 advection tensors, $\mC$, which requires $O(N^3)$ operations
and memory references on each step.  The remainder of the terms are $O(N^2)$ or
less. Unfortunately, $O(N^3)$ is a very steep cost and prohibits practical
consideration of, say, $N>200$. In this paper, our goal is to mitigate the cost with the CP decomposition. 
\end{remark}

\subsection{CP decomposition with ALS} 
\label{subsection:cpd}

The CANDECOMP/PARAFAC (CP) tensor decomposition  \cite{harshman1970foundations} 
factorizes a tensor into a sum of component rank-one tensors. For example, given
a third-order tensor $\mX \in \mathbb{R}^{s_1 \times s_2 \times s_3}$, its CP decomposition is denoted by 
\begin{equation} 
\mX \approx \llbracket A^{(1)}, A^{(2)}
   , A^{(3)} \rrbracket
\equiv \sum^R_{r=1} 
   \ua^{(1)}_r
   \circ \ua^{(2)}_r \circ \ua^{(3)}_r. \label{equation:cpd}
\end{equation}
where $\llbracket ~\rrbracket$ is the Kruskal operator
\cite{kolda2006multilinear}, which provides a shorthand notation for the sum of the
outer products of the columns of a set of matrices.  $A^{(i)} = \left[
\ua^{(i)}_1, \cdots, \ua^{(i)}_R \right] \in \mathbb{R}^{s_i \times R}$ for
$i=1,2,3$ are the factor matrices and refer to the combination of the vectors
from the rank-one components, and $R$ is the CP rank.  It is noteworthy to point
out that solving \cref{equation:cpd} with $R=1$ is already NP-hard
\cite{hillar2013most}. 

The rank of a tensor $\mX$ is defined as the smallest number of rank-one tensors that generate $\mX$ as their sum \cite{kruskal1977three}. One of the major differences between matrix and tensor rank \cite{kolda2009tensor} is that determining the rank of a specific tensor is NP-hard \cite{haastad1990tensor}. Consequently, the first issue that arises in computing a CP decomposition is how to choose the CP rank, $R$. In this paper, we consider the most common strategy by simply fitting multiple
CP decompositions with different values of $R$ until one is ``good''. 

For a given $R$ value, finding a CP decomposition for $\mX$ requires solving a nonlinear least-squares optimization problem:
\begin{equation} \min_{A^{(1)},A^{(2)},A^{(3)}}  
   \|
   \mathcal{X} -\llbracket A^{(1)}, A^{(2)}, A^{(3)} \rrbracket 
   \|_F,
   \label{equation:cp_nls}
\end{equation}
where $\|\cdot\|_F$ is the Frobenius norm and the Frobenius norm of a tensor 
is the square root of the sum of the
squares of all its elements. There are many algorithms to compute a CP decomposition. In this paper, we
consider the alternating least squares (ALS) method \cite{carroll1970analysis,
harshman1970foundations}. The idea behind the ALS is that we solve for each factor in turn, leaving all the other factors fixed. In each iteration, three subproblems are solved in sequence:
\begin{equation}
\begin{aligned}
   \min_{B\in \mathbb{R}^{s_1 \times R}}
   \|\mX - \llbracket B, A^{(2)}, A^{(3)} \rrbracket\|_F, \\
   \min_{B\in \mathbb{R}^{s_2 \times R}}
   \|\mX - \llbracket A^{(1)}, B, A^{(3)} \rrbracket\|_F, \\
   \min_{B\in \mathbb{R}^{s_3 \times R}}
   \|\mX - \llbracket A^{(1)}, A^{(2)}, B \rrbracket\|_F.
\end{aligned}
\label{equation:cp_als}
\end{equation}
Each subproblem corresponds to a linear least squares problem and is often
solved via the normal equation \cite{kolda2009tensor}, which involves tensor
contraction to solve the system of linear equations and costs $\mO(s^3R)$ if $s_i
= s$ for all $i \in \{ 1, 2, 3\}$.

\subsection{CP decomposition for partially symmetric tensor} 
    \label{subsection:cpd_sym}
    
Symmetric tensor plays an important role in many fields, for example,
chemometrics, psychometrics, econometrics, image processing, biomedical signal
processing, etc
\cite{comon2008symmetric,de2000multilinear,purvis1982full}. 
A tensor is symmetric if its elements remain the same under any permutation of
the indices \cite{kolda2009tensor}. Tensors can be partially symmetric in two or
more modes as well. 

The CP decomposition for partially symmetric tensors has been studied in
\cite{carroll1970analysis,kolda2015numerical}.  Assume the tensor $\mX$ is
symmetric in the mode-$1$ and mode-$3$, its CP decomposition is symmetric if:
\begin{equation}
\mX \approx \llbracket A^{(1)}, A^{(2)}, A^{(3)} \rrbracket = \llbracket A^{(3)}, A^{(2)}, A^{(1)} \rrbracket.
\label{equation:sym_condition}
\end{equation}
Typically, one can ensure \cref{equation:sym_condition} by enforcing the factor matrix $A^{(1)}$ and $A^{(3)}$ to be the same.
For fully symmetric tensors, setting $A^{(1)}=A^{(2)}=A^{(3)}$ results in a decomposition whose rank is referred to as the symmetric rank of a tensor.
For such tensors, the symmetric rank is often the same as the CP rank~\cite{comon2008symmetric}.

The factor $A^{(2)}$ is computed as in the ALS. For $A^{(1)}$ and $A^{(3)}$, 
we use an iterative algorithm \cite{hummel2017low} similar to the Babylonian square root algorithm, which updates $A^{(1)}$ and $A^{(3)}$ using a series of subiterations.
At each subiteration, a linear least squares problem is solved for $A^{(1)}$ with $A^{(3)}$ fixed, 
\begin{equation}
A^{(1)}_\text{next} =
\underset{B}{\mathrm{argmin}} \|\mX -
\llbracket B, A^{(2)}, A^{(1)}\rrbracket\|.
\end{equation}
Then both the first and third factors are updated with momentum, $A^{(1)}_\text{new} = \lambda A^{(1)} + (1-\lambda)A^{(1)}_\text{next}$.
\section{Accelerating G-ROM with CP decomposition}
\label{section:acc-cpd-rom}

\begin{figure}
    \centering
    \includegraphics[width=\textwidth]{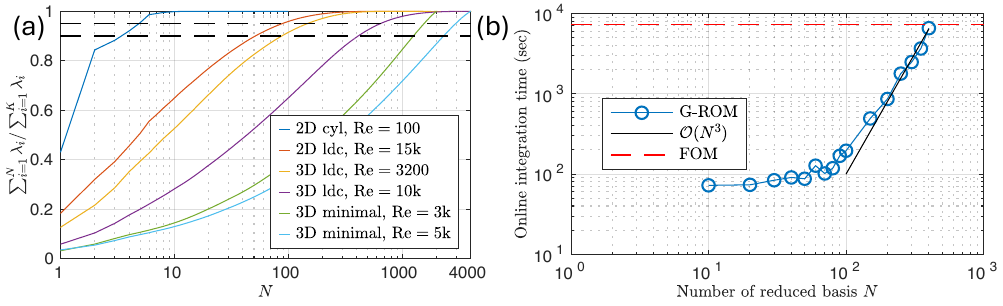}
    \vspace{-0.8cm}
    \caption{
    (a): Behavior of the accumulated POD eigenvalues for 2D
    periodic flow to 3D turbulent flows. ``cyl'', ``ldc'', and ``minimal'' refer to flow past a cylinder, lid-driven cavity, and minimal flow unit, respectively.
    (b): Behavior of the online solve time of the 
    G-ROM as a function of $N$.} 
    \label{fig:cp_motive}
\end{figure}

The motivation for considering the CP decomposition in the context of the G-ROM is illustrated in
Fig.~\ref{fig:cp_motive}. Fig.~\ref{fig:cp_motive}(a) depicts the behavior of the accumulated POD
eigenvalue for 2D periodic to 3D turbulent flows. The results show that as the
problems become advection dominated, at least $N > 100$ POD modes are needed to
capture $90\%$ of the total energy. Additionally, Fig.~\ref{fig:cp_motive}(b) shows 
the solve time for the reduced system (\ref{equation:nse_d1}) for $500$ CTUs. The results indicate that the efficiency of the ROM diminishes as $N$ increases,
due to the $\mO(N^3)$ tensor contraction cost.

The rest of this section is organized as follows: In Section \ref{subsection:cpd_rom}, we present G-ROM with the CP decomposition (CPD-ROM). In Section \ref{subsection:skew_symm}, we show the ROM tensor is partially skew-symmetric with divergence-free and certain boundary conditions. In Section \ref{subsection:cpd_skew}, we present the CP decomposition for partially skew-symmetric tensor. Finally, in Section \ref{subsection:cp_structure}, we show the ROM tensor has an underlying CP structure.

\subsection{CPD-ROM}
\label{subsection:cpd_rom}
To address the $\mathcal{O}(N^3)$ bottleneck, we consider the CP decomposition to
approximate the convection tensor $\mC$ (\ref{equation:Cu}):
\begin{equation}
   \mC_{ikj} \approx \tmC_{ikj} = \sum^R_{r=1} a_{ir}b_{kr}c_{jr}. \label{equation:approx_tensor}
\end{equation}
With (\ref{equation:approx_tensor}), the tensor contraction $\mC(\uu)\uu$ is
approximated by three matrix-vector multiplications\footnote{The first two
comes from contracting the factor matrices $B$ and $C$ with vector ${\uu}$, the
third is due to the contraction of the factor matrix $A$ with the vector $(B^T {\uu}
\ast C^T {\uu})$, where $\ast$ is the element-wise product.}: 
\begin{equation} 
    \left[\tmC ({\uu}) {\uu}\right]_{i}  =
   \sum^N_{j=0} \sum^N_{k=0} \sum^R_{r=1} a_{ir}b_{kr}c_{jr} {u}_k
   {u}_j = \sum^R_{r=1}a_{ir} (\sum^N_{k=0} b_{kr} {u}_k)
   (\sum^N_{j=0} c_{jr} {u}_j),
   \label{equation:tc_ap}
\end{equation}
fro all $i=1,\ldots,N$. 
This reduces the leading-order cost by a factor of $2N^3/6NR=N^2/3R$. 
Throughout the paper, CPD-ROM is referred to as the G-ROM
\cref{equation:nse_d1} with the advection tensor approximated by the CP decomposition 
(\ref{equation:approx_tensor}).

\subsection{Partially skew-symmetric ROM tensor} 
\label{subsection:skew_symm} 

In this subsection, we adapt the analysis in \cite{malm2013stabilization} and
show that the ROM tensor \cref{equation:Cu} is skew-symmetric in mode-$1$ and
mode-$3$ with appropriate boundary and divergence-free conditions.

We begin with the definition of the ROM tensor \cref{equation:Cu}. Because each
POD basis is a vector field $\bphi_i = (\varphi_{i,x}, \varphi_{i,y},
\varphi_{i,z})$, \cref{equation:Cu} can be further decomposed into three terms,
which represents the $x$-, $y$- and $z$- direction's contribution, respectively:
\begin{align}
    \mC_{ikj} & =\int_{\Omega}\bphi_i\cdot(\bphi_k\cdot\nabla)\bphi_j\,dV
    \nonumber \\ & = \int_{\Omega} \varphi_{i,x} (\bphi_k \cdot \nabla )
    \varphi_{j,x}~d\Omega + \int_{\Omega} \varphi_{i,y} (\bphi_k \cdot \nabla )
    \varphi_{j,y}~d\Omega + \int_{\Omega} \varphi_{i,z} (\bphi_k \cdot \nabla )
    \varphi_{j,z}~d\Omega. \label{equation:rom_tensor_decomp}
\end{align}
Without loss of generality, we consider the $x$-direction contribution.  With
the divergence theorem and the product rule, the following equation is derived:
\begin{align} 
\int_{\Omega} \varphi_{i,x} (\bphi_k \cdot \nabla ) \varphi_{j,x}~d\Omega
   & =
   -\int_{\Omega} \varphi_{j,x} (\bphi_k \cdot \nabla ) \varphi_{i,x}~d\Omega \nonumber \\ & -
   \int_{\Omega} \varphi_{i,x} \varphi_{j,x} (\nabla \cdot \bphi_k) ~d\Omega +
   \int_{\partial \Omega} \varphi_{i,x} \varphi_{j,x} \bphi_k \cdot \hat{n} dA. \label{equation:rom_tensor_expand}
\end{align}
where $\hat{\bn}$ is the outward unit normal on the boundary $\partial{\Omega}$. 
If the model problem has Dirichlet or periodic boundary conditions and the velocity reduced basis $\bphi_k$ 
is divergence-free, that is, $\nabla \cdot \bphi_k = 0$, the last two integrals vanish, leading to
\begin{equation}
\int_{\Omega} \varphi_{i,x} (\bphi_k \cdot \nabla ) \varphi_{j,x}~d\Omega
   =
   -\int_{\Omega} \varphi_{j,x} (\bphi_k \cdot \nabla ) \varphi_{i,x}~d\Omega. \label{equation:rom_tensor_skew}
\end{equation}
This indicates that the $x$-direction contribution is skew-symmetric in mode-$1$ and mode-$3$. 
For the $y$- and $z$-direction contributions, a similar equations of (\ref{equation:rom_tensor_expand})--(\ref{equation:rom_tensor_skew}) can be derived. Therefore, the ROM tensor \cref{equation:Cu} is skew-symmetric in mode-$1$ and mode-$3$, that is, $\mC_{ikj} = -\mC_{jki}$.
\begin{remark}
If the problem has inflow and outflow boundary conditions, the skew-symmetry no longer holds.
\end{remark}
\begin{remark}
In the case where skew-symmetry is no longer hold,
the tensor $\mC$ could always be decomposed into a skew-symmetric
tensor plus a low-rank tensor contributed from the boundaries, i.e., the third term in  (\ref{equation:rom_tensor_expand}).
\end{remark}
\begin{remark}
Even with 
appropriate boundary conditions, the ROM tensor \cref{equation:Cu} 
is not exactly partially skew-symmetric because
the divergence-free condition $\nabla \cdot \bphi_k=0$ is not satisfied exactly.
This is because the FOM snapshots 
only satisfy the divergence-free condition in the weak form, and is enforced to a certain accuracy ($10^{-6}$--$10^{-8}$) in practice. Therefore, small divergence errors remain. 
In this case, we enforce the partially skew-symmetry by setting $\mC_{ijk} = 0.5 (\mC_{ijk}-\mC_{kji})$ for all $i,j,k=1,\ldots,N$. 
\end{remark}

\subsection{CP decomposition for partially skew-symmetric tensor}
\label{subsection:cpd_skew}

Tensors with skew-symmetric properties appear in many fields, for example, solid-state
physics \cite{itin2022decomposition}, fluid dynamics
\cite{malm2013stabilization}, and quantum chemistry
\cite{lowdin1955quantum,szalay2015tensor}. A tensor is
skew-symmetric if its elements alternate sign under any permutation of the
indices \cite{solomonik2014massively}. Additionally, tensors can be partially skew-symmetric
in two or more modes as well, as demonstrated in the previous section.

Tensor decomposition for skew-symmetric tensors has been widely studied. For instance, see
\cite{kovac2017structure} for the Tucker decomposition, and
\cite{bader2007temporal,lundy2003application}
for decomposition into directional components (DEDICOM) and its applications. In contrast, the CP decomposition for skew-symmetric tensors has not been
widely studied, in fact, we could only find one related work
\cite{begovic2022cp} but with a different approach to enforcing skew-symmetry.

Assume the tensor $\mX$ is skew-symmetric in mode-$1$ and mode-$3$, its CP decomposition is skew-symmetric if it satisfies:
\begin{equation}
\mX \approx \llbracket A^{(1)}, A^{(2)}, A^{(3)} \rrbracket = \llbracket -A^{(3)}, A^{(2)}, A^{(1)} \rrbracket.
\label{equation:skewsym_condition}
\end{equation}
We impose the following conditions on the CP rank and factor matrices so that \cref{equation:skewsym_condition} is satisfied:
\begin{equation} 
\begin{aligned}
& \text{CP rank $R$ has to be even,} \\ 
& A^{(1)} \equiv \left[ A^{(1)}_1~ A^{(1)}_2\right],~ A^{(2)} \equiv \left[
      A^{(2)}_1~ A^{(2)}_1\right],~ A^{(3)} \equiv \left[ A^{(1)}_2~ -A^{(1)}_1 \right],
\end{aligned}
\label{equation:skew_abc_struct} 
\end{equation}
where $A^{(1)}_1$ and $A^{(1)}_2$ are matrices of size $N \times R/2$, and
the factor matrix $A^{(2)}$ has $N/2$ redundant columns. With \cref{equation:skew_abc_struct}, the CP decomposition can be expressed as:
\begin{align}
\mX \approx \llbracket A^{(1)}, A^{(2)}, A^{(3)}  \rrbracket & = \llbracket A^{(1)}_1, A^{(2)}_1, A^{(1)}_2 \rrbracket + \llbracket A^{(1)}_2, A^{(2)}_1, -A^{(1)}_1 \rrbracket \\ & = \llbracket -A^{(3)}, A^{(2)}, A^{(1)} \rrbracket,
\end{align}
and is skew-symmetric in mode-$1$ and mode-$3$.

The factor $A^{(2)}$ is computed as in the ALS.
For $A^{(1)}$ and $A^{(3)}$, we use an iterative algorithm \cite{hummel2017low}
similar to the Babylonian square root algorithm, which updates $A^{(1)}$ using a
series of subiterations.  At each subiteration, a linear least squares problem
is solved for $A^{(1)}$ with $A^{(3)}$ fixed, 
\begin{equation}
A^{(1)}_\text{next} =
\underset{B}{\mathrm{argmin}} \|\mX -
\llbracket B, A^{(2)}, A^{(3)}
\rrbracket\|_F.
\label{equation:cp_als_quad}
\end{equation} 
Then both the first and third factors are updated with momentum,
$A^{(1)}_\text{new} = \lambda A^{(1)} + (1-\lambda)A^{(1)}_\text{next}$, and
$A^{(3)}_\text{new} = \begin{bmatrix} A^{(1)}_{\text{new},2} &
-A^{(1)}_{\text{new},1} \end{bmatrix}$.

\subsection{Underlying CP structure in the ROM tensor}
\label{subsection:cp_structure}
In this section, we demonstrate that the CP decomposition is a reasonable low-rank
approximation for the advection tensor $\mC$ by showing
there is an underlying CP structure.
To illustrate this, we further expand the tensor  
$\mC$ (\ref{equation:Cu}) by substituting $\bphi_i,~\bphi_j$ and $\bphi_k$ with
$\bphi_* = (\varphi_{*,x},\varphi_{*,y})$,  
\begin{align}
    \mC_{ikj} & =\int_{\Omega}\bphi_i\cdot(\bphi_k\cdot\nabla)\bphi_j\,d\Omega
    = \int_{\Omega} \left( \varphi_{i,x} (\bphi_k \cdot \nabla )
    \varphi_{j,x}
    + \varphi_{i,y} (\bphi_k \cdot \nabla )
    \varphi_{j,y}\right)~d\Omega \nonumber \nonumber \\ & = 
    \int^1_{-1} \int^1_{-1}
    \left( \varphi_{i,x} \varphi_{k,x} \pp{\varphi_{j,x}}{x} + \varphi_{i,x} \varphi_{k,y} \pp{\varphi_{j,x}}{y} 
    \right) ~dxdy\nonumber  \\ & + 
    \int^1_{-1} \int^1_{-1}
    \left( 
    \varphi_{i,y} \varphi_{k,x} \pp{\varphi_{j,y}}{x} + \varphi_{i,y} \varphi_{k,y} \pp{\varphi_{j,y}}{y}\right) ~dxdy.
    .
    \label{equation:rom_tensor_decomp_more}
\end{align}
For simplicity, we assume the problem is two-dimensional, and the domain is simply $\Omega = [-1,~1]^2$ with one spectral element. For the general form with deformed geometry and multiple elements, we suggest \cite{lan2023development} for a 
detailed review.

(\ref{equation:rom_tensor_decomp_more}) is used to compute each component of the
tensor $\mC_{ikj}$ and the 
integration is computed using
Gaussian quadrature: \begin{align} 
   & \int^1_{-1} \int^1_{-1} 
   \varphi_{i,x} \varphi_{k,x} \pp{\varphi_{j,x}}{x} ~dxdy 
   \label{equation:conv_x_quad}
   \\  & \simeq
   \sum^{q+1}_{m=1} \sum^{q+1}_{n=1} \omega_m 
   \omega_{n}\varphi_{i,x}(\xi_{m},\xi_{n})
   \varphi_{k,x}(\xi_{m},\xi_{n}) \pp{\varphi_{j,x}(\xi_{m},\xi_{n})}{x},
   \label{equation:conv_x_quad_gll}
\end{align} 
where $\{\omega_i\}^{q+1}_{i=1}$ and $\{\xi_i\}^{q+1}_{i=1}$ denote the
Gauss-Lobatto-Legendre (GLL) quadrature weights and points, respectively.  We note that, with
$q+1$ GLL quadrature weights and points, the approximated integration is exact
if the integrand is a polynomial of at most $2q-1$. However, the integrand in (\ref{equation:conv_x_quad}) has a polynomial order larger than $2q-1$. Usually the
exactness of (\ref{equation:conv_x_quad}) is enforced by interpolating the
polynomial function onto a finer polynomial space of order $M=3q/2$, denoted as
dealiasing \cite{malm2013stabilization}.  Dealiasing increases the cost for
evaluating (\ref{equation:conv_x_quad}) but the representation stays the same
because eventually it is projected back to the
original polynomial space of order $q$. 

(\ref{equation:conv_x_quad_gll}) suggests there is an underlying CP structure:
\begin{align} 
   \sum^{q+1}_{m=1} \sum^{q+1}_{n=1} \omega_m 
   \omega_{n}\varphi_{i,x}(\xi_{m},\xi_{n}) \varphi_{k,x}(\xi_{m},\xi_{n})
   \pp{\varphi_{j,x}(\xi_{m},\xi_{n})}{x} 
   =
   \sum^{(q+1)^2}_{l=1} \rho_l a_{i,l} b_{k,l} c_{j,l}
   \label{equation:conv_x_final},
\end{align}
where
\begin{align}
   \rho_l &:= \omega_m \omega_n,\quad
   a_{i,l} :=
   \varphi_{i,x}(\xi_{m},\xi_{n}),\\ 
   b_{k,l} & :=
   \varphi_{k,x}(\xi_{m},\xi_{n}),\quad 
   c_{j,l} :=
   \pp{\varphi_{j,x}(\xi_{m},\xi_{n})}{x},
\end{align}
with the index $l := m+(n-1)(q+1)$. The analysis can be applied to the other three terms in \cref{equation:rom_tensor_decomp_more}, and a similar form of (\ref{equation:conv_x_final}) can be derived. 
(\ref{equation:conv_x_final}) suggests a theoretical bound of the CP rank 
of the advection tensor to be $4E(q+1)^2$ and $9E(q+1)^2$ for 2D and 3D problems, respectively, with $E$ being the number of spectral elements.

\section{Numerical results}
\label{section:numerical-results}

In this section, we present numerical results for the CPD-ROM introduced in
Section \ref{subsection:cpd_rom}.  For comparison purposes, we also present the
results for the G-ROM (\ref{equation:nse_d1}) and FOM. The ROMs are constructed through an offline-online procedure: In the offline
phase, the FOM is solved using the open-source code Nek5000/RS
\cite{fischer2022nekrs,fischer2008nek5000}. The POD basis and reduced
operators (\ref{equation:Cu}--\ref{equation:zeroth_contribution}) are then
constructed using NekROM \cite{Kaneko_NekROM_2025}. In addition, the CP decomposition of the tensor $\mC$ is also performed in the offline phase.  In the online phase, the reduced
system (\ref{equation:nse_d1}) is formed by loading the reduced operators and CP factor matrices and then solved using MATLAB.
We acknowledge that there are high-performance tools for tensor decomposition with
various optimization methods, for example, \cite{solomonik2014massively}. 
However, this paper's primary focus is applying the tensor decomposition with
ROM. Hence, we implement the ALS (\ref{equation:cp_als}) and ALS for skew-symmetry 
tensor (\ref{equation:cp_als_quad}), which is referred to as the ALS-skew, in Matlab for easy
use and investigation with ROM.  We use the National Center for Supercomputing
Applications (NCSA) Delta and Argonne Leadership Computing Facility (ALCF)
Polaris for the offline stage and a workstation with Intel Xeon E5-2620 CPU with
two threads for the CP decomposition and solving the ROMs. 
 
From the numerical investigation, we would like to address the following questions:
\begin{itemize}
    \item 
    We examine whether CPD-ROM provides memory savings in terms of the compression ratio and speed-up, without compromising much accuracy.
    \item Does preserving the skew-symmetry property in the approximated tensor further improve the performance of CPD-ROM?
    \item Does the advection tensor $\mC$ have low-rank structure, and how does it depends on the number of POD modes $N$, the Reynolds number $\rm Re$, and the norm used to construct the POD basis? 
\end{itemize} 
The compression ratio (CR) for the CP decomposition using ALS (\ref{equation:cp_als}) and ALS-skew
(\ref{equation:cp_als_quad}) is defined as the size ratio between the original tensor and CP models:
\begin{align} 
\tCRALS = \frac{N^3}{3NR} = \frac{N^2}{3R},\quad \tCRALSS =
\frac{N^3}{\frac{3}{2}NR} = \frac{2N^2}{3R}\label{eq:compress_ratio}.
\end{align}
A factor of $2$ in $\tCRALSS$ arises from the redundant columns
in the factor matrix $A^{(2)}$ and the skew-symmetric structure in $A^{(1)}$ and
$A^{(3)}$ (Section \ref{subsection:cpd_skew}).  These ratios do not account for the skew-symmetry of the original tensor, as it is difficult to exploit in
storage or tensor contraction. We note that the cost reduction in the tensor contraction remains the same for both ALS and ALS-skew, 
namely $N^2/3R$, because both CP models require three matrix-vector multiplications (\ref{equation:tc_ap}) to approximate the operation. Unless otherwise noted, CR refers to $\tCRALSS$. 

The rest of this section is organized as follows.
In Section \ref{subsection:grom-vs-cpd-rom}, we compare the CPD-ROM introduced in Section \ref{subsection:cpd_rom} with the G-ROM (\ref{equation:nse_d1}) across four test problems, from 2D periodic flow to 3D turbulent flows.
We use the $H^1_0$-based POD basis functions for both ROMs, motivated by the low-rank structure
observed under the $H^1_0$ norm in Section \ref{subsubsection:low-rank}.
In Section \ref{subsection:full-vs-core}, 
we investigate the performance of CPD-ROM with the full tensor approximated and with the core tensor approximated.
In Section \ref{subsection:als-vs-als-quad}, we investigate whether preserving skew-symmetry in the CP decomposition improves CPD-ROM performance.
In Section \ref{subsubsection:low-rank}, we investigate the low-rank structure of the advection tensor formed using the $L^2$- and $H^1_0$-POD basis functions using the SVD. 
In Section~\ref{subsubsection:svd_vs_cpd}, we compare the low-rank approximations obtained via SVD and CP decomposition in terms of the relative residual and compression ratio. We also compare the performance of the CPD-ROM with the SVD-ROM, where the advection tensor is approximated using SVD.

\subsection{Performance comparison of CPD-ROM and G-ROM}
\label{subsection:grom-vs-cpd-rom}

\subsubsection{2D flow past a cylinder} 
\label{subsub:cylinder}

Our first example is the 2D flow past a cylinder at the Reynolds number $\rm
Re=100$, which is a canonical test case for ROMs due to its robust and
low-dimensional attractor, manifesting as a von Karman vortex street for $\rm Re
= UD/\nu > 34.37$ \cite{ding2021free}, where
$D$ is the cylinder diameter and $U$ is the free-stream velocity. The computational domain is $\Omega = [-2.5 \, D,17 \, D] \times [-5 \, D,5 \, D]$, with $D=1$ and the cylinder centered at $[0,0]$.

The reduced basis functions  $\{\bphi_i\}^N_{i=1}$ 
are constructed 
via POD from 
$K=100$
snapshots collected over $100$ convective time units $(D/U)$, after von Karman
vortex street is developed. 
The zeroth mode $\bphi_0$ is set to be the mean velocity over this window. 
The initial condition for the G-ROM and CPD-ROM is obtained by projecting
the lifted snapshot at $t=500$ (in convective time units) onto the reduced
space. We use $N=20$ POD basis functions, which capture over $99\%$ of the snapshot energy.

We test the G-ROM and CPD-ROM in the reproduction regime, defined as the time interval in which the snapshots were collected, and consider the total drag in the $x$-direction as the quantity of interest (QOI). The total drag on the cylinder is defined as:
\begin{align}
\boldsymbol{F}_D = \oint_{\Gamma} (-\nu \nabla \bu + p) d{\boldsymbol{A}},
\label{equation:total_drag}
\end{align}
with $\Gamma$ being the surface of the cylinder. 
We refer to \cite{kaneko2022augmented} for computing the pressure drag in the
ROM without solving the pressure solution.

\begin{figure}
    \centering
    \includegraphics[width=1\columnwidth]{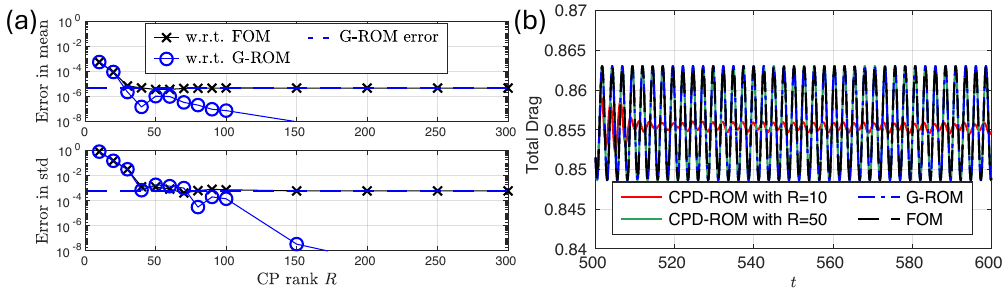}
    \vspace{-0.8cm}
    \caption{2D flow past a cylinder at $\rm Re=100$: (a) Relative errors in the
    mean and standard deviation of the total drag in $x$-direction for the
    CPD-ROM as a function of $R$, computed relative to the FOM and G-ROM. 
    G-ROM error relative to the FOM is shown as a blue dashed line.
    (b) The total drag in the $x$-direction of the FOM, G-ROM, and CPD-ROM.}
    \label{fig:2dfpc}
\end{figure}

Fig.~\ref{fig:2dfpc}(a) shows the relative errors in the mean and standard
deviation of the total drag in the $x$-direction for the CPD-ROM as a
function of the CP rank $R$. 
These errors are calculated with respect to both the
FOM and G-ROM for comparison.  Additionally, a horizontal dashed line
represents the relative error of the G-ROM calculated with respect to the FOM, which is independent of $R$.
We found that the errors of the CPD-ROM with respect to the G-ROM decrease as $R$ increases.
When $R \ge 200$, the errors in both the mean and standard deviation fall below 
$10^{-8}$, indicating that the CPD-ROM converges to the G-ROM. The errors of the CPD-ROM with respect to the FOM initially decrease as $R$ increases and eventually reach the same level of accuracy as the G-ROM (represented by the blue dashed line).
This behavior is expected, as the G-ROM serves as the reference model for the CPD-ROM.

Fig.~\ref{fig:2dfpc}(b) shows the history of the total drag in
$x$-direction for the CPD-ROM with $R=10$ and $R=50$, along with the FOM and G-ROM
results.  
Despite the poor approximation to the advection tensor made by the CP decomposition with 
$R=10$, the total drag in the CPD-ROM does not blow up but remains stable. This behavior could be attributed to the model problem having a robust and
low-dimensional attractor. With $R=50$, the total drag in the CPD-ROM shows good
agreement with the results from both the FOM and G-ROM.

\subsubsection{2D lid-driven cavity (2D LDC)}
\label{subsub:2dldc}

Our next example is the 2D LDC problem at $\rm Re=\num{15000}$,
which is a more challenging test case than the 2D flow past a
cylinder. As demonstrated in \cite{fick2018stabilized}, this problem requires
$N\ge 60$ POD modes for the G-ROM to accurately reconstruct solutions and QOIs.
A detailed description of the FOM setup for this problem can be found in
our previous work \cite{kaneko2020towards}. 

The reduced basis functions $\{\bphi_i\}^N_{i=1}$ are constructed
via POD from $K=\num{2000}$ snapshots collected in the statistically steady state region
over 
$[\num{6000.4},~ \num{6200}]$ with a sampling time of $\Delta t_s=0.1$.
The zeroth mode $\bphi_0$ is 
the mean velocity over this window. The initial condition for the G-ROM and CPD-ROM is  
obtained by projecting the lifted snapshot at $t=\num{6000}$ (in convective time
units) onto the reduced space. We use $N=200$ for both models. This value is chosen to ensure that G-ROM remains accurate relative to the FOM.
We test both the G-ROM and CPD-ROM in the reproduction regime and consider the energy and the fluctuated velocity energy as the QOIs, which are defined as 
\begin{equation}
    \mathrm{E}(t)=\frac{1}{2}\int_{\Omega}\|\bu(x,t)\|^2_2,\quad
    \mathrm{E}(t)_{\tfluc}=\frac{1}{2}\int_{\Omega}\|\bu(x,t)-\langle\bu\rangle(x)\|^2_2,
    \label{equation:ene_fluc} 
\end{equation} where $\|\cdot\|_2$ is the Euclidean
norm\footnote{$\mathrm{E}(t)_{\tfluc}$ in (\ref{equation:ene_fluc}) is usually
referred to as the turbulent kinetic energy, however, for 2D and 3D low Reynolds
number flow, there is no turbulence, therefore, we refer
(\ref{equation:ene_fluc}) as the fluctuated velocity energy for not
confusing the reader.}. 
These QOIs are also considered in the 3D lid-driven cavity and the 3D minimal flow unit problems.

\begin{figure}
    \centering
    \includegraphics[width=1\columnwidth]{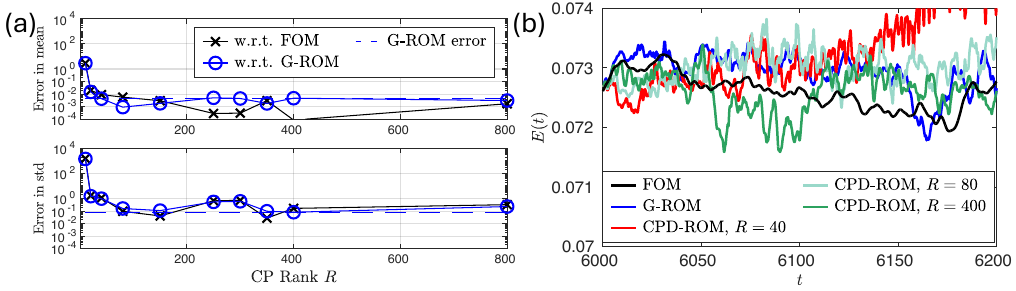}
    \vspace{-0.8cm}
    \caption{2D LDC at $\rm Re=\num{15000}$: (a) Relative errors in the mean and standard deviation of the energy 
    for the CPD-ROM as a function of $R$, computed relative to the FOM and G-ROM. G-ROM error relative to the FOM is shown as a blue dashed line. 
    (b) The energy of the FOM, G-ROM, and CPD-ROM.} 
    \label{fig:2dldc_ene}
\end{figure}

Fig.~\ref{fig:2dldc_ene}(a) shows the relative errors in the mean and standard deviation of the energy $E(t)$ for the CPD-ROM as a function of
the CP rank $R$. 
These errors are calculated with respect to both the FOM and 
G-ROM for comparison. Additionally, a horizontal dashed line indicates 
the relative error of the G-ROM with respect to the FOM, which is independent of $R$. 
Unlike the results for the 2D flow past a cylinder, where the errors relative to the G-ROM decrease to machine precision as $R$ increases, we observed that, in this test case, the errors initially
decrease 
but then begin to fluctuate around the G-ROM's
accuracy level ($0.8\%$ in mean and $10\%$ in the standard deviation) once $R\ge 80$.
This behavior is expected because the model problem is chaotic, and even with a small approximation error, it can lead to a different solution trajectory.
In addition, we found that the errors with respect to the FOM decrease as $R$ increases and eventually reach the same level of accuracy as the G-ROM,
similar to what was 
observed for the 2D flow past a cylinder.  Furthermore, the error in the mean becomes smaller than that of the G-ROM for $R \ge 250$.

We further plot the history of $\energy$ for the CPD-ROM with $R=40,~80$, and $400$ in
Fig.~\ref{fig:2dldc_ene}(b), along with the results for the FOM and G-ROM.
With $R=40$, the reconstructed $\energy$ is not accurate. However, with $R=80$ and $R=400$, the results 
are comparable to those from the FOM and G-ROM. We note that $\energy$ of the CPD-ROM with $R=400$ initially
agrees with the G-ROM but diverges onto a different trajectory at $t \approx \num{6010}$.
This is expected due to the approximation error made in the CP decomposition. 

\begin{figure}
    \centering
    \includegraphics[width=1\columnwidth]{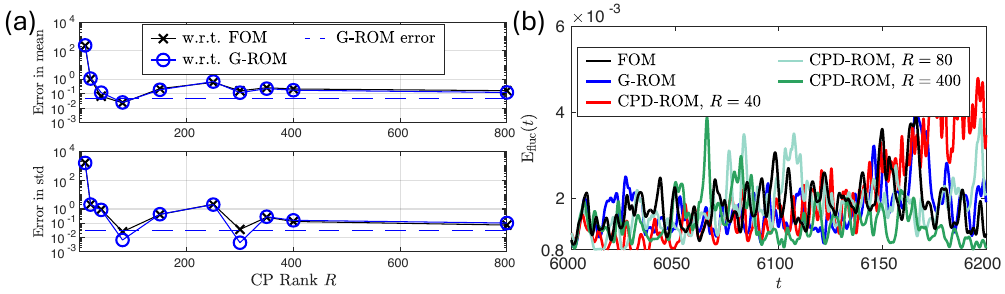}
    \vspace{-0.8cm}
    \caption{2D LDC at $\rm Re=\num{15000}$: (a) Relative errors in the mean and standard deviation of the fluctuated velocity energy for the CPD-ROM as a function of $R$, computed relative to the FOM and G-ROM. 
    G-ROM error relative to the FOM is shown as a blue dashed line.
    (b) The fluctuated velocity energy of the FOM, G-ROM, and CPD-ROM.}
    \label{fig:2dldc_intke}
\end{figure}

Fig.~\ref{fig:2dldc_intke}(a) shows the relative errors in the mean and 
standard deviation of the fluctuated velocity energy $\efluc$ for the CPD-ROM as a function of the CP rank $R$. These errors are calculated with respect to both the FOM and G-ROM for comparison.
Additionally, a horizontal dashed line shows the relative error of the G-ROM
calculated with respect to the FOM quantity, which is independent of $R$.
We observed a similar error behavior as in $\energy$ results, but with generally higher
magnitudes. This is expected because $\efluc$ is a more challenging QOI than $\energy$. In addition, 
the error of the CPD-ROM with respect to the FOM is generally larger than
the G-ROM's error.  We note that the error increases when
$R$ increases 
from $80$ to $300$. We suspect this behavior is due to the random initial condition in the CP decomposition. To investigate this, we further ran five additional trials of the CP decomposition for $R=80,~150,~250$ and $300$. We did not find a small error at $R=80$ as in Fig.~\ref{fig:2dldc_intke}(a); however, we did find that the error decreases in general as $R$ increases. We further plot the history of $\efluc$ 
for the CPD-ROM with $R=40,~80$, and
$400$ in Fig.~\ref{fig:2dldc_intke}(b), along with the results for the FOM
and G-ROM. As in the case of $\energy$, $\efluc$ at $R = 40$ is not accurate, while at $R = 80$ and $R = 400$, $\efluc$ is comparable to those from the FOM and the G-ROM.

In summary, in terms of $\energy$, the CPD-ROM with $R=80$ 
achieves the same
level of accuracy as the G-ROM for both the mean and standard deviation. 
For $\efluc$, the CPD-ROM with $R=300$ achieves accuracy similar to the G-ROM for the standard deviation, but the error in the mean is four times greater.
Both $R$ values offer significant reductions in the compression ratio (CR) (\ref{eq:compress_ratio}) and cost of evaluating the nonlinear term (\ref{equation:tc_ap}). Specifically, 
with $R=80$, CR is $333.3$ and the cost is reduced by a factor of $166.6$.
With $R=300$, CR is $88.9$, leading to a cost reduction by a factor of $44.4$.
 
For $\num{200000}$ time steps, the G-ROM with $N=200$ takes 
about $\num{1167}$ seconds to solve, with the tensor contraction kernel occupying about
$86.9\%$ of the time. In contrast, 
the CPD-ROM with $R=300$ takes 
about $101$ seconds, and the CP kernel
(\ref{equation:tc_ap}) occupies only $14.5\%$ of this time. 
Despite a theoretical speedup of $6.63$ from a $44.4$-fold cost reduction, we observe an actual speedup of $11.5$. The discrepancy is due to the favorable cache effect. The machine 
has L1, L2, and L3 caches of $0.384$ MB, $3$ MB, and $30$ MB, respectively. With $N=200$, the advection tensor $\mC$ is of size $210$ MB, whereas the total
size of the factor matrices with $R=300$ is $4.5$ MB only, resulting in additional space in the L3 cache to store the ROM operators. All results are from single-threaded MATLAB runs to ensure a fair comparison, since MATLAB optimizes the G-ROM for large 
$N$ but provides no such optimization for the CPD-ROM.

\subsubsection{3D lid-driven cavity (3D LDC)}

We next consider the non-regularized 3D LDC problem at $\rm Re=\num{10000}$, which poses a greater challenge for the G-ROM than the 2D cases discussed above. Results for $\rm Re=\num{3200}$ are qualitatively similar and can be found in the dissertation \cite{tsai2023pmordevelopment}.
Following \cite{kaneko2022augmented}, the FOM mesh consists of a
tensor-product array of $E = 16^3$ elements with a Chebyshev distribution. With polynomial order $q=7$, this yields $\cN \approx 2$ million unknowns.  

The reduced basis functions $\{\bphi_i\}^N_{i=1}$ are constructed via POD from $K=\num{4000}$ snapshots collected in the statistically steady state region over $[\num{2725.125},~ \num{3225}]$ with a sampling time of $\Delta t_s=0.125$.
The zeroth mode $\bphi_0$ is the mean velocity 
over this window.  The initial condition for the G-ROM and CPD-ROM is obtained by
projecting the lifted snapshot at $t=\num{2725}$ onto the reduced space. We use $N=400$ for both models. This value is chosen to ensure that G-ROM remains accurate relative to the FOM.

\begin{figure}
    \centering
    \includegraphics[width=1\columnwidth]{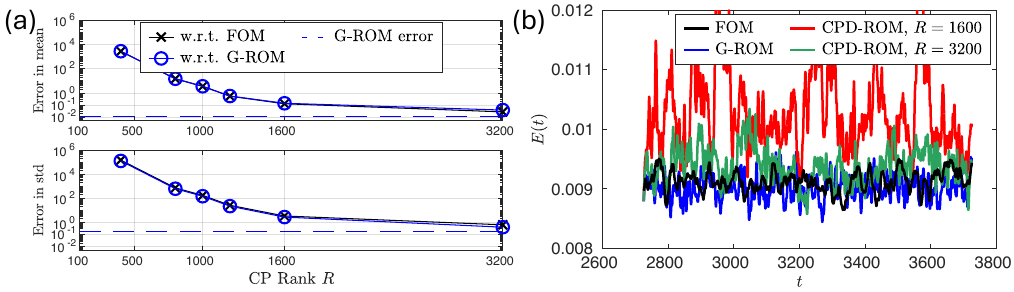}
    \caption{3D LDC at $\rm Re=\num{10000}$: (a) Relative errors in 
    the mean and standard deviation of the energy for the CPD-ROM as a function of $R$, computed relative to the FOM and G-ROM. G-ROM error relative to the FOM is shown as a blue dashed line.
    (b) The energy 
    of the FOM, G-ROM, and CPD-ROM.}
    \label{fig:3dldc_ene}
\end{figure}

Fig.~\ref{fig:3dldc_ene}(a) shows the relative errors in the mean and standard deviation of the energy $\energy$ (\ref{equation:ene_fluc}) for the CPD-ROM as a function of the CP rank $R$. 
The errors are computed with respect to both the FOM and G-ROM for comparison.  Additionally, a horizontal dashed line indicates the relative error of  
the G-ROM with respect to the FOM, which is independent of  
$R$.
Unlike the previous two test cases, G-ROM and CPD-ROM are evaluated in a  
predictive regime over $[\num{2725},~\num{3725}]$, extending $500$ CTUs beyond the training interval.
We observed that the errors with respect to both the FOM and G-ROM decrease as $R$
increases.  Specifically, we found that at least $R \ge \num{1600}$ is required for the
CPD-ROM to predict the mean energy with an error of $10\%$, and $R = \num{3200}$ is needed to predict a
reasonable standard deviation. 
Fig.~\ref{fig:3dldc_ene}(b) shows
the history of $\energy$ for the CPD-ROM with $R=\num{1600}$ and $R=\num{3200}$, along with the results for the FOM and
the G-ROM.  We found that the result with $R=\num{1600}$ shows a much larger fluctuations than the
FOM and G-ROM, and an improvement is observed with $R=\num{3200}$. 

\begin{figure}
    \centering
    \includegraphics[width=1\columnwidth]{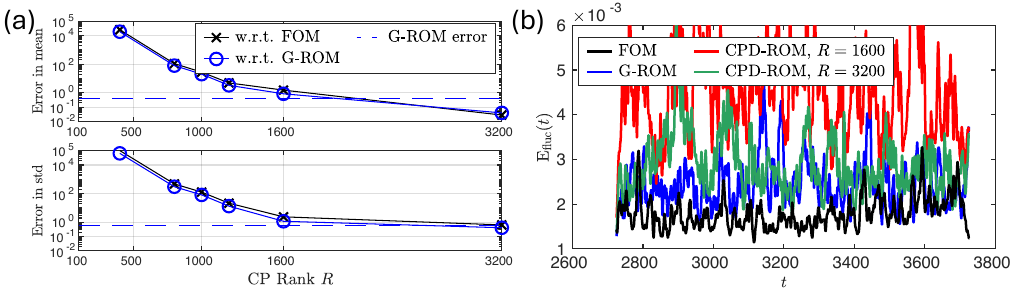}
    \vspace{-0.7cm}
    \caption{3D LDC at $\rm Re=\num{10000}$: (a) Relative errors in the mean and standard deviation of the fluctuated velocity energy for the CPD-ROM as a function of 
    $R$, computed relative to the FOM and G-ROM.  
    G-ROM error relative to the FOM is shown as a blue dashed line. (b) The fluctuated velocity energy of the FOM, G-ROM, and CPD-ROM.} 
    \label{fig:3dldc_intke}
\end{figure}

Fig.~\ref{fig:3dldc_intke}(a) shows the relative errors in the mean and 
standard deviation of the fluctuated velocity energy $\efluc$ (\ref{equation:ene_fluc}) for the CPD-ROM  as a function of the CP rank $R$.  We observed a similar error behavior as in the energy results for the CPD-ROM.  In addition, we found that the errors for
both the CPD-ROM and G-ROM are larger than those associated with the energy. 
Furthermore, with $R=\num{3200}$, the CPD-ROM matches the G-ROM's accuracy in predicting the standard deviation and outperforms the G-ROM in predicting the mean.
Fig.~\ref{fig:3dldc_intke}(b) shows the history of $\efluc$ for the CPD-ROM with $R=\num{1600}$
and $\num{3200}$, along with the FOM and G-ROM results. The result with $R=\num{3200}$ is similar to those of the G-ROM and FOM, while the result with $R=\num{1600}$ has a larger fluctuation.

In summary, to predict the mean energy within $10\%$ error, 
CPD-ROM requires $R=\num{1600}$, giving a compression ratio of $66.7$, and the cost is reduced by a factor of $33.3$. For the mean fluctuated
velocity energy, a larger CP rank 
$R=\num{3200}$ is required, with 
a compression ratio of $33.3$ and the cost is reduced by a factor of $16.7$. Predicting the standard deviation of both QOIs within 
$10\%$ error requires
$R \ge \num{3200}$. 

For $\num{1000000}$ time steps, the G-ROM with $N=400$ takes about $\num{43819}$ seconds, with the tensor contraction kernel occupying about
$97\%$ of the solve time. In contrast, the 
the CPD-ROM with $R=\num{3200}$ takes about $\num{4269}$ seconds, with the CP kernel
(\ref{equation:tc_ap}) occupying only $64.8\%$ of the time.  This yields a speedup by 
a factor of $10.3$, 
which is close to the
theoretical value of $11.4$. We do not get a larger speedup as in the previous test case because
the size of the factor matrices 
is $91$ MB, which exceeds 
the L3 cache size. 

\subsubsection{The minimal flow unit (MFU)}
We next consider 
the MFU at $\rm Re=\num{5000}$, which presents strong turbulent features
while maintaining simplified flow dynamics, resulting in significantly lower
computational costs compared to a full channel flow simulation
\cite{jimenez1991minimal}. Results for $\rm Re=\num{3000}$ are qualitatively similar and can be found in the dissertation \cite{tsai2023pmordevelopment}. Following the setup in
\cite{jimenez1991minimal}, the streamwise and spanwise lengths of the
channel are set to $0.6 \pi h$ and $0.18\pi h$, respectively. The channel
half-height is set to $h = 1$. 
The FOM mesh consists of an array of $8 \times 18 \times 4$
elements in the $x \times y \times z$ directions. With polynomial order $q=9$, this yields $\cN \approx 420$ thousands unknowns. 

The reduced basis functions $\{\bphi_i\}^N_{i=1}$ are constructed via 
POD from $K=\num{4000}$ snapshots collected in the statistically steady state region over $[\num{3000.125},~ \num{3500}]$ with a sampling time of $\Delta t_s=0.125$. The
zeroth mode $\bphi_0$ is set to the mean velocity over this window. The initial condition for the G-ROM and CPD-ROM is 
obtained by projecting the lifted snapshot at $t=\num{3000}$ onto the reduced space. 
We use $N=400$ for both the G-ROM and CPD-ROM and evaluate both models in a predictive regime over
$[\num{3000},~\num{4000}]$, extending $500$ CTUs beyond the training interval. 

\begin{figure}
    \centering
    \includegraphics[width=1\columnwidth]{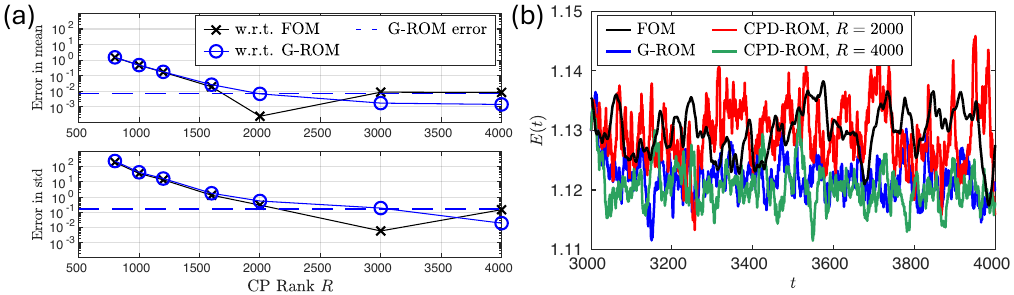}
    \vspace{-0.7cm}
    \caption{MFU at $\rm Re =\num{5000}$: 
    (a) Relative errors in the mean and standard deviation of the energy for the CPD-ROM as functions of $R$, computed with respect to the FOM and G-ROM. G-ROM error with respect to the FOM is shown as a blue dashed line. (b) The energy of the FOM, G-ROM, and CPD-ROM.} 
    \label{fig:3dmfu_energy}
\end{figure}

Fig.~\ref{fig:3dmfu_energy}(a) shows the relative errors in the
mean and standard deviation of the energy $\energy$ (\ref{equation:ene_fluc}) for the CPD-ROM as a function of the CP rank $R$.
These errors are calculated with respect to both the FOM and G-ROM for
comparison.  Additionally, a horizontal dashed line indicates the relative error of  
the G-ROM with respect to the FOM, which is independent of 
$R$.
We found that at least $R \ge \num{1200}$ is required for the CPD-ROM to predict the mean
with an error of $10\%$. To predict a reasonable standard deviation,
a higher rank of $R=\num{3000}$ is required.  In addition, the errors relative to
the G-ROM decrease as $R$ increases, whereas the errors relative to the FOM
begin to fluctuate around the G-ROM accuracy level once $R\ge \num{2000}$. 
The history of $\energy$ of the CPD-ROM with $R=\num{2000}$ and $R=\num{4000}$ is shown in
Fig.~\ref{fig:3dmfu_energy}(b), along with the results for the FOM
and G-ROM. We observed that the result with $R=\num{4000}$ closely matches that of the G-ROM, but is slightly underestimated compared to the FOM. In contrast, the result with $R=\num{2000}$ shows larger fluctuations.

\begin{figure}[ht]
    \centering
    \includegraphics[width=1\columnwidth]{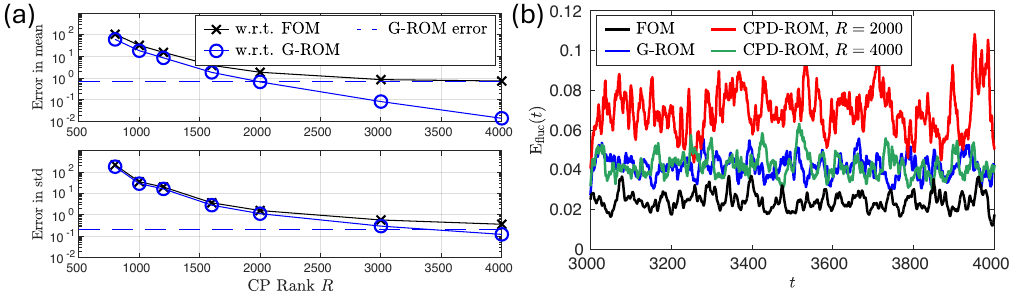}
    \vspace{-0.7cm}
    \caption{MFU at $\rm Re =\num{5000}$: 
    (a) Relative errors in the mean and standard deviation of the fluctuated velocity energy for the CPD-ROM as functions of $R$, computed relative to the FOM and G-ROM. G-ROM error relative to the FOM is shown as a blue dashed line. (b) The fluctuated velocity energy of the FOM, G-ROM, and CPD-ROM.} 
    \label{fig:3dmfu_intke}
\end{figure}

Fig.~\ref{fig:3dmfu_intke}(a) shows the relative errors in the mean and
the standard deviation of the fluctuated velocity energy $\efluc$  (\ref{equation:ene_fluc}) for the CPD-ROM 
as a function of the CP rank $R$.
As $R$ increases, the errors with respect to the FOM decrease and eventually approach the accuracy level of the G-ROM.  However, these errors remain large because the G-ROM itself is not accurate, with errors of
$70\%$ in the mean and $21\%$ in the standard deviation.  To improve
the results, $N>400$ is required.
Fig.~\ref{fig:3dmfu_intke}(b) shows
the history of $\efluc$ for the CPD-ROM with $R=\num{2000}$ and $R=\num{4000}$, along with the results for the
FOM and G-ROM.  We found that the result with $R=\num{2000}$ is not accurate, and the
result with $R=\num{4000}$ is similar to the result of the G-ROM. However, both
are inaccurate compared to the FOM. 

In summary, to predict the mean and 
standard deviation of $\energy$ within $10\%$ error requires $R=\num{3000}$ in the CPD-ROM, yielding a compression ratio of $35.5$, and the cost is reduced by 
a factor of $17.7$. For $\efluc$, despite the CPD-ROM with $R=\num{4000}$ achieves the same level of accuracy as the G-ROM, the errors relative to
the FOM remain large (over $20\%$).  This is due to the inaccuracy of the G-ROM. To
resolve this issue, one must consider $N > 400$. 

For $\num{400000}$ time steps, the G-ROM with $N=400$ takes about $\num{17470}$ seconds to solve, with the tensor contraction kernel occupying about
$97\%$ of the solve time. In contrast, the CPD-ROM with $R=\num{3000}$ takes only $\num{1559}$ seconds, with the CP kernel
(\ref{equation:tc_ap}) occupying only $64.1\%$ of the time.  
This yields a speedup by a factor of $11.2$, which is close to the theoretical value of $11.8$.

\subsection{Performance of CPD-ROM with approximated full and core tensors}
\label{subsection:full-vs-core}

Recall the definition of the ROM advection tensor:
\begin{equation}
    \mC_{ikj} = \int_{\Omega} \bphi_i \cdot (\bphi_k \cdot \nabla) \bphi_j~dV.
\end{equation}
The full tensor includes the contributions from the zeroth mode $\bphi_0$ and is
defined as $\mC_{ikj}$ with $i=1,\ldots,N$ and $j,k=0,\ldots,N$. 
The core tensor excludes contributions from the zeroth mode and is defined as
$\mC_{ikj}$ with $i,j,k=1,\ldots,N$. 

In this section, we investigate the performance of CPD-ROM with the full tensor approximated (CPD-ROM-Full) and with the core tensor approximated (CPD-ROM-Core). We expect the CPD-ROM-Core to outperform the CPD-ROM-Full, because the contributions from the zeroth mode (i.e., the $C_1$ and $C_2$ matrices defined in (\ref{equation:zeroth_contribution})) are not approximated. 
For fair comparison, ALS (\ref{equation:cp_als}) is used to compute the CP decomposition for both
tensors.

\begin{figure}
    \centering
    \includegraphics[width=1\columnwidth]{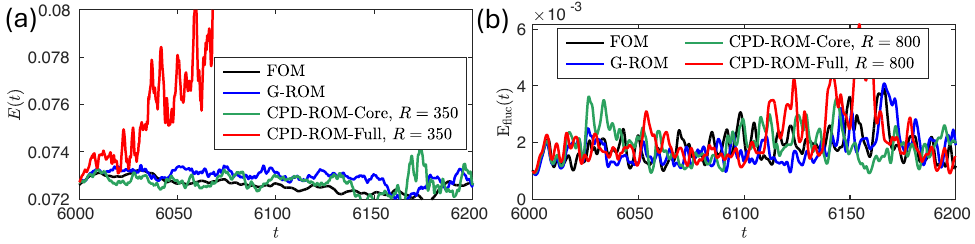}
    \vspace{-0.7cm}
    \caption{2D LDC at $\rm Re=\num{15000}$: 
    (a) The energy and (b) the fluctuated velocity energy 
    of the FOM, G-ROM, and CPD-ROM with approximated full and core tensors.}
    \label{fig:2dldc_full_core_compare}
\end{figure}

Fig.~\ref{fig:2dldc_full_core_compare} compares the performance of the CPD-ROM-Full and CPD-ROM-Core for the 2D LDC at $\rm Re=\num{15000}$. 
Fig.~\ref{fig:2dldc_full_core_compare}(a) shows the history of the energy $\energy$ for both models with $R=350$, along with the FOM and G-ROM results. 
At $R=350$, CPD-ROM-Core is able to reproduce the energy, 
with errors relative to the FOM 
below $0.1\%$ for the mean and
around $27\%$ for the standard deviation. In contrast, CPD-ROM-Full is not
stable, and its solution diverges after $t>6030$. 

We further assess the approximation error in the full and core tensors made by CP decomposition. Given a tensor $\mC$ and its approximated tensor $\wmC$, the relative residual is defined as:
\begin{align}
\relres := \frac{\|\mC - \wmC\|_F}{\|\mC\|_F}, \label{equation:relative_residual}
\end{align} where $\|\cdot \|_F$ is the Frobenius norm. 
Despite finding a significant performance difference between the CPD-ROM-Core
and CPD-ROM-Full in Fig.~\ref{fig:2dldc_full_core_compare}(a), 
the relative residual (\ref{equation:relative_residual}) of the approximated tensor in both cases is similar.  Note that the relative residual of the approximated core tensor is
measured against the full tensor instead of the core tensor.

Fig.~\ref{fig:2dldc_full_core_compare}(b) shows the history of the fluctuated velocity energy $\efluc$ 
for both models with $R=800$, along with the FOM and G-ROM results.
A larger rank, $R=800$, is required for the CPD-ROM-Core to achieve good accuracy because $\efluc$ is generally a much more challenging quantity of interest compared to $\energy$.
Consistent with the $\energy$ results, the CPD-ROM-Core is more stable and achieves better accuracy than the 
CPD-ROM-Full.
Despite the chaotic nature of the problem, $\mathrm{E}_\tfluc$ of the CPD-ROM-Core behaves
similarly to the FOM, with relative errors of $4\%$ in both the mean and standard deviation. 
In contrast, $\mathrm{E}_\tfluc$ of the CPD-ROM-Full overshoots in later
time, with relative errors of $16\%$ in the mean and $71\%$ in the standard
deviation.

We next investigate the performance of the CPD-ROM-Full and CPD-ROM-Core for the
3D lid-driven cavity at $\rm Re=\num{3200}$ in both the reproduction
and the predictive regimes.
\begin{figure}[!ht]
    \centering
    \includegraphics[width=1\columnwidth]{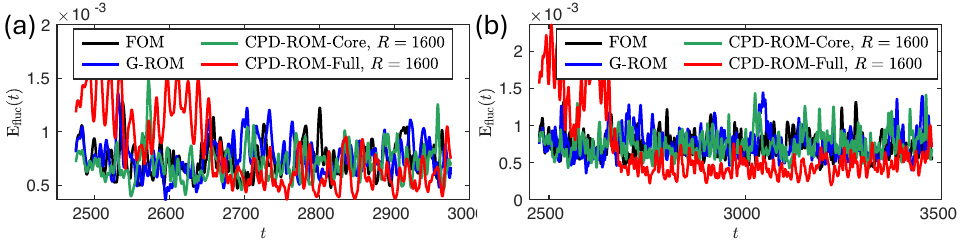}
    \vspace{-0.8cm}
    \caption{3D LDC at $\rm Re=\num{3200}$: 
    The fluctuated velocity energy of the FOM, G-ROM, and CPD-ROM with approximated full and core tensors in (a) the reproduction and (b) the predictive regimes.}
    \label{fig:3dldc_re3200_full_core_compare}
\end{figure}
Fig.~\ref{fig:3dldc_re3200_full_core_compare}(a) shows the history of $\efluc$ for the two models in the reproduction regime with $R=\num{1600}$, along
with the FOM and G-ROM results. 
The CPD-ROM-Core reproduces $\efluc$ with relative errors of $5\%$ in the mean and $0.4\%$ in the standard
deviation compared to the FOM.
In contrast, although the CPD-ROM-Full solution remains stable, it
is less accurate, with relative errors of $9\%$ in the mean and $127\%$ in the standard
deviation.

Fig.~\ref{fig:3dldc_re3200_full_core_compare}(b) shows the results in the predictive regime.
The CPD-ROM-Core accurately predicts $\efluc$ with errors of $2\%$ in the
mean and $5\%$ in the standard deviation. In contrast, the CPD-ROM-Full is not
accurate, with errors of $10\%$ in the mean and $200\%$ in the standard
deviation.

\subsection{Performance of CPD-ROM with skew-symmetry preserved}
\label{subsection:als-vs-als-quad}

In this section, we investigate whether preserving the skew-symmetry of the
approximated core tensor $\wmC$ improves the performance of the CPD-ROM.
We consider three test problems: 2D LDC, 3D LDC, and MFU. For each test case, we compare the CPD-ROM with skew-symmetry preserved
(CPD-ROM-Skew) to the standard CPD-ROM without this property.
These test problems are chosen specifically because their boundary conditions satisfy the requirement for the ROM tensor to be skew-symmetric
(\ref{equation:rom_tensor_skew}).
In addition, for each test problem, we enforce the tensor $\mC$ to be
skew-symmetric, that is, $\mC_{ijk} = 0.5 (\mC_{ijk}-\mC_{kji})$ for all
$i,j,k=1,\ldots,N$, as discussed in Section \ref{subsection:skew_symm}.
To compute a CP decomposition that preserves this skew-symmetry, we employ the ALS-skew algorithm introduced in Section~\ref{subsection:cpd_skew}. 

\begin{figure}
    \centering
    \includegraphics[width=1\columnwidth]{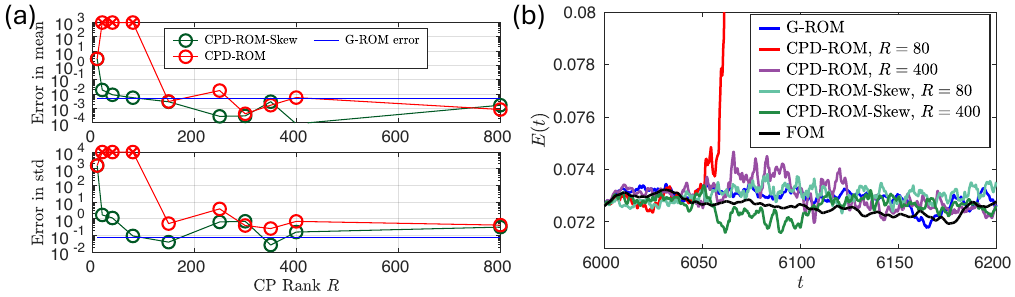}
    \vspace{-0.7cm}
    \caption{
    2D LDC at $\rm Re=\num{15000}$: 
    (a) Relative error in the mean and standard deviation of the energy for the CPD-ROM and CPD-ROM-Skew as a function of $R$.
    G-ROM error relative to the FOM is illustrated as a blue line.
    Circle markers with crosses indicate NaN values. 
    (b) 
    The energy of the FOM, G-ROM, CPD-ROM-Skew, and CPD-ROM.} 
    \label{fig:2dldc_skew_plain}
\end{figure}

Fig.~\ref{fig:2dldc_skew_plain} presents a performance comparison between the CPD-ROM-Skew and CPD-ROM for the 2D LDC at 
$\rm Re=\num{15000}$ in terms of the energy $\energy$.
Fig.~\ref{fig:2dldc_skew_plain}(a) shows the relative errors in the mean 
and standard deviation of $\energy$ 
for both models as a function of CP rank $R$. 
G-ROM errors, which are independent of $R$, are included as horizontal lines. Values of $R$ that cause NaN errors in the CPD-ROM due to instability are indicated with a circle and cross. The results show that the CPD-ROM-Skew remains stable and achieves good accuracy at low $R$ compared to the CPD-ROM.  However, as $R$ increases, both models behave similarly. 
This behavior is expected because skew-symmetry addresses stability, not
accuracy.  This observation is consistent with the behavior of full-order
models, where recovery of skew symmetry (e.g., through dealiasing) is known to
provide stability but does not, in general, improve accuracy.  This point was
discussed in numerous early works by Orszag and co-authors (e.g.,
\cite{fox1973pseudospectral,orszag1972comparison,orszag1974numerical}).

Fig.~\ref{fig:2dldc_skew_plain}(b) shows the history of $\energy$ for both models with $R=80$ and $R=400$, along with the results for the FOM and G-ROM. 
At $R=80$, we observed that $\energy$ of the CPD-ROM deviates from the FOM. In contrast, $\energy$ of the CPD-ROM-Skew remains stable, with errors less than $1\%$ in the mean and approximately $10\%$ in the standard deviation. 
At $R=400$, the approximated tensor in both models is sufficiently accurate so to ensure the stability of the CPD-ROM. 

\begin{figure}[ht]
    \centering
    \includegraphics[width=1\columnwidth]{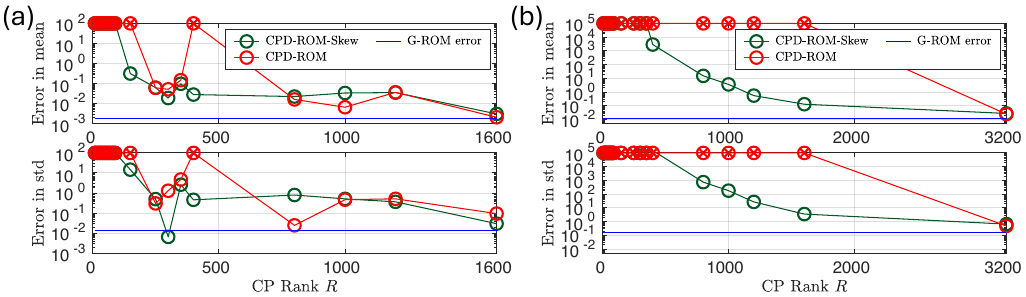}
    \vspace{-0.7cm}
    \caption{
    3D LDC: 
    Relative errors in the mean and
    standard deviation of the energy for the CPD-ROM-Skew and CPD-ROM as a
    function of $R$ for (a) $\rm Re=\num{3200}$ and (b) $\rm Re=\num{10000}$. 
    G-ROM error relative to the FOM is shown as a blue line.
    Circle markers with crosses indicate NaN values .} 
    \label{fig:3dldc_skew_plain}
\end{figure}

The comparison is extended to the 3D LDC at $\rm Re=\num{3200}$ and $\rm Re=\num{10000}$, with results shown in Fig.~\ref{fig:3dldc_skew_plain}.
Figs.~\ref{fig:3dldc_skew_plain}(a)--(b) show
the relative errors in the mean and standard deviation of $\energy$ for the
CPD-ROM-Skew and CPD-ROM 
as a function of CP rank $R$ for $\rm
Re=\num{3200}$ and $\rm Re=\num{10000}$, respectively. 
For the CPD-ROM, we found that the error yields NaN values at several small $R$ values for $\rm Re=\num{3200}$, and at all considered $R$ values except $R=\num{3200}$ for $\rm Re=\num{10000}$. These NaNs occur because the solution diverges due to instability. 
Similar to the 2D LDC, we found that the CPD-ROM-Skew remains stable and outperforms the CPD-ROM at small $R$ values, while both models behave similarly as $R$ increases.

\begin{figure}[!ht]
    \centering
    \includegraphics[width=1\columnwidth]{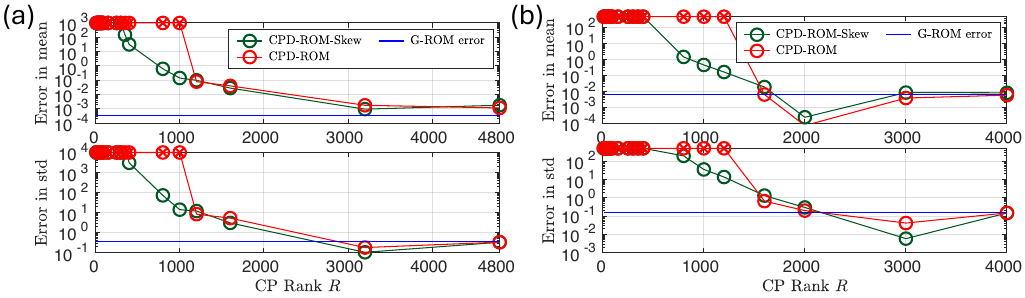}
    \vspace{-0.7cm}
    \caption{
    MFU: 
    Relative errors in the mean and standard
    deviation of the energy for the CPD-ROM-Skew and CPD-ROM as a function of
    $R$ for (a) $\rm Re=\num{3000}$ and (b) $\rm Re=\num{5000}$.  
    G-ROM error relative to the FOM is shown as a blue dashed line. Circle markers with crosses indicate NaN values.}
    \label{fig:3dminimal_skew_plain}
\end{figure}

Fig.\ref{fig:3dminimal_skew_plain} extends the comparison to the MFU at $\rm Re=\num{3000}$ and $\rm Re=\num{5000}$.
Figs.\ref{fig:3dminimal_skew_plain}(a)--(b) show the relative errors in the mean and standard deviation of $\energy$ for both models as functions of CP rank $R$ for $\rm Re=\num{3000}$ and $\rm Re=\num{5000}$, respectively. Consistent with the 2D and 3D LDC results, CPD-ROM-Skew remains stable and outperforms CPD-ROM at low $R$, while both models behave similarly as $R$ increases.
Moreover, the 3D LDC and MFU results 
indicate that CPD-ROM requires a larger 
$R$ to remain stable at higher Reynolds numbers, 
while CPD-ROM-Skew remains stable at relatively smaller $R$.
 
\subsection{Numerical investigation with singular value decomposition}
\label{subsection:svd-vs-cpd}

\subsubsection{Low-rank structure ablation} 
\label{subsubsection:low-rank}

We investigate the low-rank structure of two tensors $\mC$, one formed using the $L^2$-POD basis functions and the other using the $H^1_0$-POD basis functions, by analyzing the normalized singular values of their mode-$1$, mode-$2$ and mode-$3$ matricizations \cite{kolda2009tensor}, $C_{(1)},~C_{(2)}$ and $C_{(3)}$. We note that if the tensor $\mC$ is partially skew-symmetric (\ref{equation:rom_tensor_skew}), each row of $C_{(3)}$ preserves such skew symmetry.

\begin{figure}
    \centering
    \includegraphics[width=1\columnwidth]{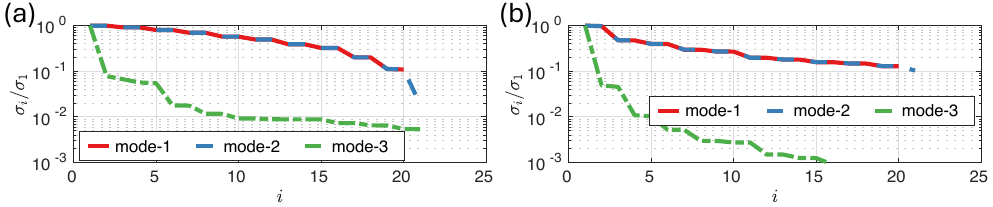}
    \vspace{-0.7cm}
    \caption{2D flow past a cylinder at $\rm Re=100$: Behavior of the singular
    values of the mode-1, mode-2, and mode-3 matricizations of the tensor $\mC$
    formed with (a) $L^2$-POD and (b) $H^1_0$-POD basis functions.}
    \label{fig:2dfpc_svd}
\end{figure}

Fig.~\ref{fig:2dfpc_svd} shows the behavior of the singular values of $C_{(1)}$,
$C_{(2)}$ and $C_{(3)}$ for the tensor $\mC$ formed using the $L^2$- and
$H^1_0$-POD basis functions in the 2D flow past a cylinder at $\rm Re=100$.
The singular values of $C_{(3)}$ decay much faster than those  
of $C_{(1)}$ and $C_{(2)}$ for both basis types, 
indicating that there is a low-rank structure in $C_{(3)}$ in comparison to $C_{(1)}$ and $C_{(2)}$. In addition, the singular values of all three matricized matrices formed using the $H^1_0$-POD basis decay much faster than those formed using the $L^2$-POD basis. As highlighted in
\cite{fick2018stabilized}, the $H^1_0$-POD basis functions are expected to perform
better in capturing small-scale structures and distinguishing them from
large-scale structures in the solution, compared to the $L^2$-POD basis function.
Therefore, it is not surprising that the singular value of all three matrices
decay faster with the $H^1_0$-POD basis. 

\begin{figure}[!ht]
    \centering
    \includegraphics[width=1\columnwidth]{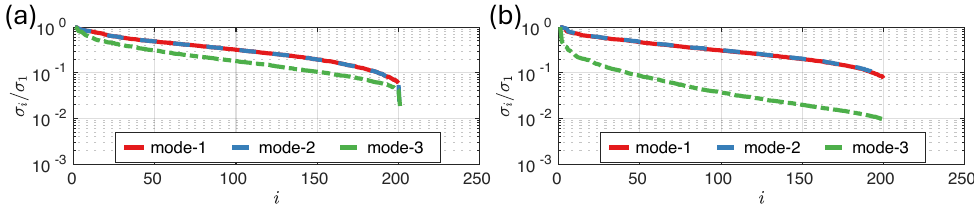}
    \vspace{-0.7cm}
    \caption{2D LDC at $\rm Re=\num{15000}$: Behavior of the singular
    values of the mode-1, mode-2, and mode-3 matricizations of the tensor $\mC$
    formed with (a) $L^2$-POD and (b) $H^1_0$-POD basis functions.} 
    \label{fig:2dldc_svd}
\end{figure}

Fig.~\ref{fig:2dldc_svd} shows the behavior of the singular values of $C_{(1)}$,
$C_{(2)}$ and $C_{(3)}$ for the tensor $\mC$ formed using the $L^2$- and $H^1_0$-POD basis
functions in the 2D lid-driven cavity at $\rm Re=\num{15000}$. 
We found that the singular values of $C_{(1)}$ and $C_{(2)}$ behave similarly 
for both basis types, while the singular values of $C_{(3)}$ decay much
faster. 
In contrast to the 2D flow past a cylinder case, the decay associated with the $L^2$-POD basis 
is much slower. This discrepancy suggests 
the presence of a low-rank structure in $C_{(3)}$ associated with the $H^1_0$-POD basis. 

Similar investigations were carried out for the 3D LDC and the MFU, yielding
consistent results with those observed in the 2D LDC. 
These results show that the matrix $C_{(3)}$ formed using the $H^1_0$-POD basis exhibits a low-rank structure. 

\begin{figure}[ht]
    \centering
    \includegraphics[width=1\columnwidth]{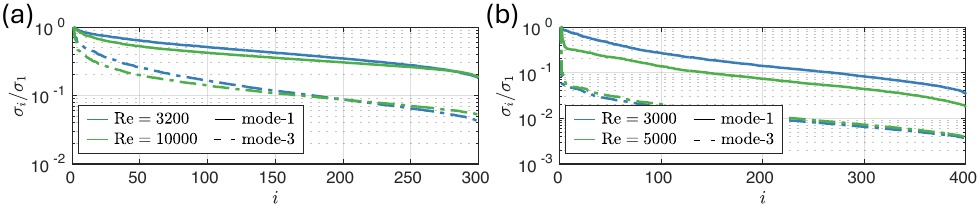}
    \vspace{-0.7cm}
    \caption{Effect of Reynolds number on the behavior of the singular values of
    mode-1 and mode-3 matricizations of the tensor $\mC$ formed using $H^1_0$-POD basis functions in
    (a) the 3D LDC and (b) the MFU.} 
    \label{fig:3d_h10_svd}
\end{figure}

We further investigate the impact of the Reynolds number on the behavior of
singular values for the 3D LDC and the MFU in Fig.~\ref{fig:3d_h10_svd} with a
primary focus on the tensor $\mC$ formed using the $H^1_0$-POD basis functions.  
Fig.~\ref{fig:3d_h10_svd}(a) shows the behavior of the singular values of
$C_{(1)}$ and $C_{(3)}$ for the 3D LDC at $\rm Re=\num{3200}$ and $\rm Re=\num{10000}$.  We
found that the singular values of both $C_{(1)}$ and $C_{(3)}$ decay slightly
faster at $\rm Re=\num{10000}$. 
Fig.~\ref{fig:3d_h10_svd}(b) shows
the behavior of the singular values of $C_{(1)}$ and $C_{(3)}$ for the MFU at
$\rm Re=\num{3000}$ and $\rm Re=\num{5000}$. A faster decay in the
singular values of $C_{(1)}$ is observed at $\rm Re=\num{5000}$, while the decay of the singular values of $C_{(3)}$ remains similar across both Reynolds numbers. 
Overall, we didn't
find a significant difference in the behavior of singular values between low and
high Reynolds numbers. Nevertheless, a low-rank structure continues to be observed in
$C_{(3)}$ formed using the $H^1_0$-POD basis, even at higher Reynolds numbers.

\subsubsection{Performance comparison of CPD-ROM and SVD-ROM}
\label{subsubsection:svd_vs_cpd}

In this section, we first compare the low-rank approximations obtained via SVD and CP decomposition in terms of the relative residual (\ref{equation:relative_residual}) and compression ratio (\ref{eq:compress_ratio}), and compare the performance of CPD-ROM with the SVD-ROM, in which the advection tensor is approximated using SVD.
Throughout this study, we focus on the CP decomposition
with skew-symmetry preserved. 

The relative residual for the SVD is computed using
(\ref{equation:relative_residual}) but with the mode-3 matricization $C_{(3)}$
and its SVD approximation $\widehat{C}_{(3)}$. The compression ratio (CR) for the
ALS-skew and SVD is defined as
\begin{equation}
    \tCRALSS = \frac{N^3}{\frac{3}{2}NR} = \frac{2N^2}{3R},\quad \tCRSVD = \frac{N^3}{N^2R} = \frac{N}{R}. 
\end{equation}

\begin{figure}[!ht]
    \centering
    \includegraphics[width=1\columnwidth]{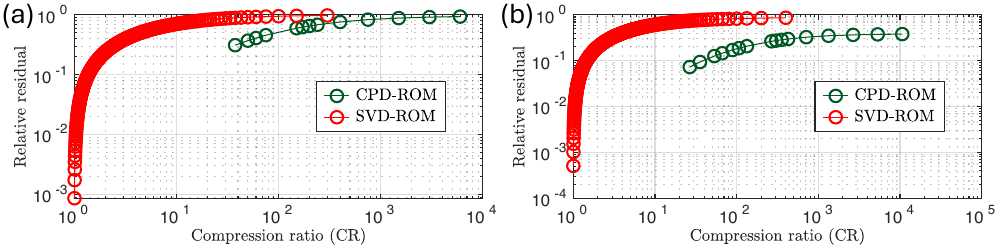}
    \vspace{-0.7cm}
    \caption{Behavior of the relative residual of the approximated tensor
    using SVD and CP decomposition as a function of CR in (a)
    the 3D LDC and (b) the MFU.} 
    \label{fig:relr_cr}
\end{figure}

Fig.~\ref{fig:relr_cr} shows the behavior of the relative residual as a function
of CR for the SVD and CP decomposition in the 3D LDC and the MFU. 
We found that the CP decomposition outperforms the SVD in terms of the compression
ratio, achieving a higher CR value for a fixed relative residual.
While the SVD can achieve lower relative residuals than the 
CP decomposition, its corresponding CR remains close to $1$, offering almost no speed-up. Moreover, as shown in previous sections, a relative residual between
$0.35$ to $0.1$ is sufficient for the CPD-ROM to achieve comparable accuracy to the
G-ROM.

We now turn to the performance comparison between the SVD-ROM and the CPD-ROM. In the SVD-ROM, the tensor contraction $\mC(\uu)\uu$ is evaluated using the
approximated mode-$3$ matricization of the tensor $\mC$:
\begin{align}
   C_{(3)} \approx \widehat{C}_{(3)} = U_R \Sigma_R V^T_R,
\end{align}
where $U_R$ and $V_R$ are the left and right singular vector matrices of size
$\mathbb{R}^{N\times R}$ and $\mathbb{R}^{N^2 \times R}$, respectively, and
$\Sigma_R$ is the singular value matrix of size $\mathbb{R}^{R \times R}$.

\begin{figure}[ht]
    \centering
    \includegraphics[width=1\columnwidth]{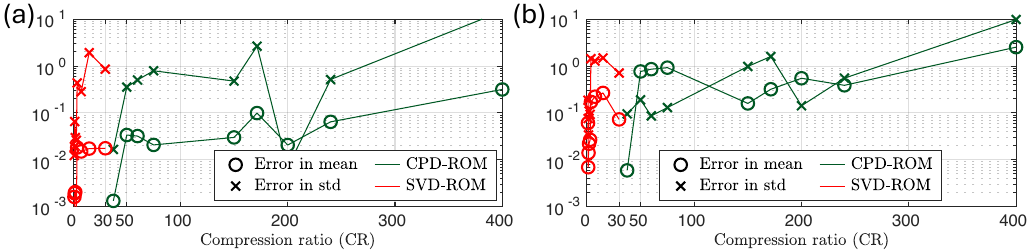}
    \vspace{-0.7cm}
    \caption{3D LDC at $\rm Re=\num{3200}$: 
    Relative error in the mean and standard deviation
    of (a) the energy and (b) the fluctuated velocity energy for the
    SVD-ROM and CPD-ROM as a function of CR.} 
    \label{fig:svd_cpd_compare_3dldc}
\end{figure}
Fig.~\ref{fig:svd_cpd_compare_3dldc} shows the
relative error in the mean and standard deviation of the energy $\energy$ and 
the fluctuated velocity energy $\efluc$ as a function of the compression ratio (CR)
for the 3D LDC. In both the mean and standard deviation of $\energy$ and $\efluc$, we found that the CPD-ROM achieves a
larger CR compared to the SVD-ROM for a given accuracy. 

\begin{figure}[!ht]
    \centering
    \includegraphics[width=1\columnwidth]{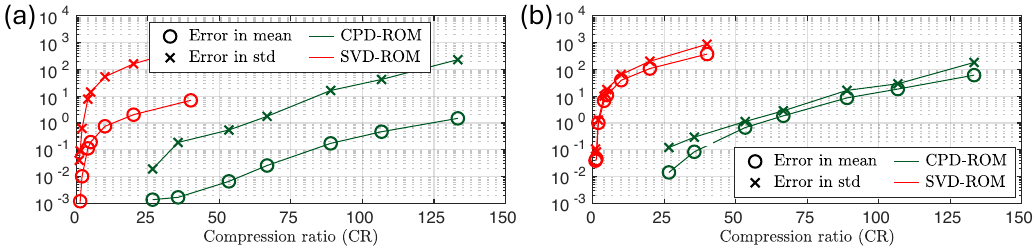}
    \vspace{-0.7cm}
    \caption{MFU at $\rm Re=\num{5000}$: 
    Relative error in the mean and standard deviation
    of (a) the energy and (b) the fluctuated velocity energy for the
    SVD-ROM and CPD-ROM as a function of CR.} 
    \label{fig:svd_cpd_compare_3dmfu}
\end{figure}

Fig.~\ref{fig:svd_cpd_compare_3dmfu} shows the
relative error in the mean and standard deviation of $\energy$ and $\efluc$ as a function of the compression ratio (CR)
for the MFU. As in the 3D LDC case, we found that the CPD-ROM achieves a
larger CR compared to the SVD-ROM for a given accuracy in both the mean and standard deviation of $\energy$ and $\efluc$. 
These results demonstrate that the CP decomposition provides a more effective low-rank approximation of the advection tensor $\mC$ than the SVD.

\section{Conclusions and discussions}
\label{sec:conclusions}

In this work, we propose a novel approach to accelerate Galerkin reduced-order models (G-ROMs) 
by leveraging the CANDECOMP/PARAFAC (CP)
decomposition to approximate the ROM  advection tensor by a sum of $R$ rank-1 tensors. 
Through numerical investigation across several 2D and 3D flow problems, we demonstrate that the resulting CPD-ROM achieves 
at least a $10$-fold speed-up, with the computational cost of evaluating the nonlinear term reduced by a factor of at least $16.7$, 
while maintaining acceptable accuracy. 
In addition, we show that the skew-symmetry
preserving CPD-ROM is more stable in both the reproduction and predictive regimes, and, for a given accuracy, enables the use of a smaller CP
rank $R$, which results in a larger speed-up.
Moreover, we compare the CPD-ROM with the SVD-ROM, in which the advection tensor is approximated using singular value decomposition (SVD). For a given accuracy in approximating the advection tensor and predicting quantities of interest, we show that the CPD-ROM outperforms the SVD-ROM in terms of compression ratio.
Finally, we demonstrate that the advection tensor formed using the $H^1_0$-POD basis functions exhibits a low-rank structure in its mode-$3$ matricization, and that this low-rank structure is preserved even at higher Reynolds numbers.

The first step in the numerical investigation of the CPD-ROM 
is encouraging.  There are, however, several other research directions that should
be pursued next.  
For example, CP decomposition can be applied to the nonlinear advection term in the energy equation, further reducing computational costs in fluid–thermal applications. More broadly, the proposed approach is applicable to other PDEs with quadratic nonlinearities, such as the Vlasov equations \cite{tsai2023accelerating}.
As discussed in Section~\ref{subsection:skew_symm}, for general flow problems, the reduced tensor can be decomposed into a
skew-symmetric part with a low-rank tensor contributed by the boundary conditions.  
It would be valuable to assess CPD-ROM performance in that setting. 
In this study, the alternating least squares (ALS) method and its variant were
employed to compute the CP decomposition.  However, ALS may exhibit slow or no
convergence, especially when high accuracy is required
\cite{singh2021comparison}. 
Therefore, exploring alternative optimization methods to construct the CPD-ROM could potentially yield more accurate CP approximation and result in a greater speed-up.
Another important direction for future work is comparing the skew-symmetry preserving CPD-ROM with other G-ROMs that employ existing hyper-reduction techniques to address the nonlinear evaluation cost.
Finally, another possibility is to explore other
tensor decompositions, such as Tucker decomposition, to assess if they offer additional advantages over CP decomposition.

\bibliographystyle{plain}
\bibliography{tensor}
\end{document}